\documentclass[journal]{IEEEtran}
\usepackage{graphicx}
\usepackage{multirow}
\graphicspath{{figures/}}
\usepackage{tikz}
\usepackage[american]{circuitikz}
\tikzset{fontscale/.style = {font=\relsize{#1}}}
\usetikzlibrary{arrows,shapes,snakes,automata,backgrounds,petri}
\usetikzlibrary{positioning,arrows}
\tikzset{main node/.style={circle,fill=blue!20,draw,minimum size=1cm,inner sep=0pt},
}
\usepackage{amssymb}
\usepackage{mathrsfs}
\usepackage{setspace}
\usepackage{MnSymbol}
\usepackage{booktabs}
\usepackage{enumerate}
\usepackage{color}
\usepackage{algcompatible}
\usepackage[justification=centering]{caption}
\usepackage[section]{algorithm}
\usepackage{algpseudocode}
\usepackage{algcompatible}

\usepackage{amsthm}
\usepackage{xfrac}

\def\figurename{\small Fig.}
\def\tablename{\small Table}

\renewcommand{\Re}{\rm Re~} 
\newcommand{\MU}{\mu} 
\newcommand{\SZP}{\xi} 
\newcommand{\sign}{{\rm sign}} 
\newcommand{\diag}{{\rm diag}}

\newcommand{\n}{\tt{n}}
\newcommand{\m}{\tt{m}}
\newcommand{\p}{\tt{p}}

\newcommand{\R}{\ensuremath{\mathbb{R}} }

\newcommand{\C}{\ensuremath{\mathbb{C}} }

\newcommand{\Rn}{\R^{\n} }

\newcommand{\Rnn}{\R^{\n \times \n} }
\newcommand{\Rnp}{\R^{\n \times \p} }
\newcommand{\Rmn}{\R^{\m \times \n} }
\newcommand{\Rmp}{\R^{\m \times \p} }
\newcommand{\B}{\ensuremath{\mathfrak{B}}}

\newcommand{\Kmax}{K_{\rm max}}
\newcommand{\Kmin}{K_{\rm min}}

\newcommand*{\Scale}[2][4]{\scalebox{#1}{$#2$}}%

\newtheorem{theo}{Theorem}[section]
\renewcommand\thesection{\arabic{section}}

\usepackage{algpseudocode}

\newtheorem{corollary}[theo]{Corollary}

\newtheorem{definition}[theo]{Definition}
\newtheorem{example}[theo]{Example}

\newtheorem{lemma}[theo]{Lemma}

\newtheorem{remark}[theo]{Remark}

\renewcommand{\baselinestretch}{0.975}

\begin{document}
\title{Interlacing properties of system-poles, system-zeros and spectral-zeros in MIMO systems}
\author{Sandeep~Kumar
\thanks{Sandeep Kumar is in ACEM, DRDO, Nasik and also in the Department
of Electrical Engineering, Indian Institute of Technology Bombay. 
Madhu Belur is in the Department
of Electrical Engineering, Indian Institute of Technology Bombay.
\mbox{Emails: \tt{ sandeepkumar.iitb@gmail.com, belur@iitb.ac.in } }}
and~Madhu~N.~Belur}


\maketitle


\begin{abstract}
 SISO passive systems with just one type of memory/storage
element (either only inductive or
only capacitative) are known to have real poles and zeros, and further,
with the zeros interlacing poles (ZIP).
Due to a variety of definitions of the notion of a system zero, and due
to other reasons described in the paper,
results involving ZIP have not been
extended to MIMO systems. This paper formulates
conditions under which MIMO systems too have interlaced poles and zeros.

This paper next focusses on the notion of a `spectral zero' of a system,
which has been well-studied in various contexts: for example,
spectral factorization, 
optimal charging/discharging of a dissipative system, and 
even model order reduction.
We formulate conditions under which the spectral zeros of a MIMO
system are real, and further,
conditions that guarantee that the system-zeros, spectral zeros and the poles
are all interlaced. 

The techniques used in the proofs involve new results in Algebraic Riccati
equations (ARE) and Hamiltonian matrices, and these results help in formulating
new notions of positive-real balancing, and inter-relations with the existing
notion of positive-real balancing; we also relate the positive-real
singular values 
with the eigenvalues of the extremal ARE solutions in the proposed
`quasi-balanced' forms.

\end{abstract}

\begin{IEEEkeywords}
RC/RL realizability, MIMO impedance/admittance transfer matrices,
real spectral zeros, zeros interlacing poles (ZIP),
spectral zeros interlacing, balancing methods,
symmetric state-space realizable systems
\end{IEEEkeywords}

\section{Introduction} \label{sec:intro}

It is well-known that SISO passive systems containing resistors and
only one type of memory/storage element, namely capacitative or inductive, have
only real poles and zeros, and further, that these are interlaced. 
In a related context, `spectral zeros' of a system is a well-studied notion:
they play a key role in model order reduction, in dissipativity studies,
spectral factorization: more about this in Section~\ref{subsec:literature}.
In the context of passive circuits, when considering the
problem of minimizing the energy required to charge
an initially-discharged circuit to a specified state vector,
and analogously that of maximizing the energy extractable by
discharging an initially charged circuit to a fully-discharged state,
the spectral zeros correspond to the exponents of the exponential
trajectories at optimum charging/discharging. A spectral zero being real
signifies that the charging/discharging profile contains
no oscillations, and thus the trajectory is purely
an exponentially increasing (while charging the circuit)
or exponentially decreasing (while discharging the circuit) profile.

This paper addresses these notions for MIMO systems and formulates
conditions under which the poles and zeros
are interlaced. A key difficulty in extending
SISO pole/zero interlacing properties to MIMO system is identifying
the right notion of a system-zero, due to the variety of (non-equivalent)
definitions of a system zero.

This paper next formulates conditions under which MIMO systems 
have real spectral zeros, and further conditions for interlacing of
system-zeros, spectral zeros and system-poles. While many of the
interlacing results are known for the SISO case only, some of
this paper's MIMO-case conclusions
turn out to follow under simpler conditions for the SISO case, and are new
results for the SISO case too. 

The techniques used in this paper involve new results
Algebraic Riccati Equation (ARE)
and Hamiltonian matrix properties: we apply these results to
the case of positive real balancing. A summary of contribution in this paper
follows later in Section~\ref{subsec:summary-contri}.

\subsection{Background and related work} \label{subsec:literature}

Systems with zeros-interlacing-poles (ZIP) have been well-studied, see,
for example, \cite{willems1976realization,liu1998model,srinivasan1997model},
and references therein. It has been shown that such systems admit symmetric state-space realization. Passive systems which admit symmetric state-space realization are part of a broader class of systems called 
relaxation systems \cite{willems1976realization}. 
These systems correspond to physical systems which have only one ``type'' of 
energy storage possibility, e.g. only potential energy or only kinetic energy,
but not both.
It has been noted that Resistor-Inductor (RL) and Resistor-Capacitor (RC)
have this property and, conversely, under mild assumptions,
ZIP systems can be realized as impedance or admittance of RC/RL systems.
In view of this, in our paper, when considering a transfer
function and its inverse, we often use $Z(s)$ and $Y(s)$ to denote a
transfer function/matrix as impedance or admittance of an underlying
passive circuit. 

Beyond the classical areas of RC/RL realization, passive systems, especially
those having the ZIP property, have received much attention in the
literature recently too: see \cite{flagg_2012}, \cite{ha_2015},
\cite{padoan_2014}, \cite{srinivasan1997model} for example.  In the
context of model order reduction.
ZIP systems also find applications in the modelling of non-laminated
axial magnetic bearings \cite{herzog_2009}, and in biological systems \cite{sandberg_2013}.
In the context of the ability to compose a system as parallel interconnection
of `simple compartments', \cite{astolfi_2004} brings out the close
link with  ZIP systems.  
In the context of Hankel singular values, \cite{mironovskii_2013} 
studies a class of linear dynamical systems, known as modally balanced systems,
in which the system-poles are proportional to its Hankel singular values: these
systems too are shown to exhibit the ZIP property.
In the context of fractional-order systems, \cite{maione_2019} utilizes
the pole-zero interlacing architecture for various applications
like synthesis of fractional order PID controllers \cite{aware_2017}
and discrete time fractional operators.

However, all papers listed above, both classic and recent, focus
only on SISO systems. Despite our best efforts in searching for interlacing 
related results in the literature on MIMO systems,
just a mention that `ZIP systems can also be defined for MIMO systems \cite{Yan_1987}'
was found in \cite{liu1998model}, notwithstanding that \cite{Yan_1987} deals with
a slightly different notion of interlacing called `even interlacing' (also
termed `parity interlacing property'), in the context of 
stabilizing a MIMO system using a stable controller.
This paper focusses on extending and formulating SISO Zero-Interlacing-Pole
(ZIP) results for the MIMO case, and lack of progress in this
direction is not very surprising since
there are examples of multi-port RC circuits having driving point
impedances with nonreal poles/zeros and,
together with mutual inductances, even nonminimum-phase
zeros (see \cite[Sec.~8.6]{seshu_1964} for these examples).
Another reason explaining the difficulty in extending
ZIP results to the MIMO case is the variety of
(non-equivalent) definitions of a system-zero for a MIMO system: see \cite{WymSai83},
\cite[Section~6.5.3]{Kai84}. 

In order to obtain ZIP results for MIMO systems, and in the context
of spectral zeros of a system being real, we use {\em symmetric state-space}
realizable systems (see Definition~\ref{def:symmetric state-space realizable} below).
Systems with such a realization, called symmetric systems,
have been well-studied: firstly, they exhibit ZIP \cite{willems1976realization},\cite{srinivasan1997model},\cite{liu1998model}.
Secondly, models of networks of systems often
naturally give rise to a symmetric state-space realization: symmetry
often coming because of a reciprocity in the interaction between 
neighbours.
Such realizations have found applications in
multi-agent networks \cite{briegel_2011},\cite{yoon_2013}. 

Later in Section~\ref{sec:examples}, we consider a multi-agent network
in the context of MIMO systems exhibiting ZIP. We first consider below a
passive circuit to relate realizability as RC or RL when ZIP property
is satisfied. This also motivates the use of $\Sigma_Y: (A_Y,B_Y,C_Y,D_Y)$
and $\Sigma_Z: (A_Z,B_Z,C_Z,D_Z)$ in the context of relating 
state-space realizations of $G(s)$ and of its inverse.

\subsection{RC/RL-networks, interlacing and spectral zeros: example} \label{subsec:network-example}
Consider a strictly passive SISO system $\Sigma$ with
transfer function 
\[
G(s)=\frac{(s+2)(s+5)}{(s+1)(s+3)}= 1 + \frac{2}{s+1}+\frac{1}{s+3}.
\]
The system-zeros $\{-2,-5\}$ interlace the system-poles $\{-1,-3\}$.
Obviously, the inverse system $\Sigma^{-1}$ defined by the
transfer function $G(s)^{-1}$ also has the ZIP property.
A network realization of this system needs only a single type of energy storage element.
The system can be realized as either  RC or RL network depending on assigning the transfer function of the system as impedance $Z(s):=G(s)$ or admittance $Y(s):=G(s)$
of the network respectively.  Though this is well-known, we motivate questions
addressed in this paper using this example.
 
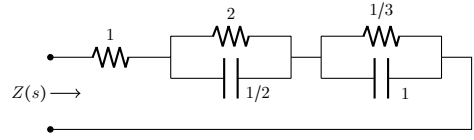
\begin{figure}[h!]
  \centering
  \ctikzset{bipoles/resistor/height=0.3}
  \ctikzset{bipoles/resistor/width=0.5}
  \begin{tikzpicture} [scale=0.8,transform shape, every node/.style={scale=0.8}]
  \draw (1,0) to[R,l=$1$]  (3,0) -- (3,0.35) to[R=$2$] (5,0.35) -- (5,0);
  \draw (5,0) -- (5.5,0);
  \draw (5.5,0) -- (5.5,0.35) to[R=$1/3$] (7.5,0.35) -- (7.5,0);
  \draw (7.5,0) --(8,0) -- (8,-1.2) -- (1,-1.2); 
  \draw (3,0) -- (3,-0.35);
  \draw  (5,-0.35) -- (5,0);
  \draw (5,-0.35) to[C] (3,-0.35);
  \node[] at (4.45,-0.6) {$1/2$};
  \draw (5.5,0) -- (5.5,-0.35);
  \draw  (7.5,-0.35) -- (7.5,0); 
  \draw (7.5,-0.35) to[C] (5.5,-0.35);
  \node[] at (6.9,-0.6) {$1$};
  \draw[->]  (1,-0.6) -- (1.5,-0.6);
  \node[] at (0.65,-0.6) {$Z(s)$};
  \node[circ] at(1,0) {}; 
  \node[circ] at(1,-1.2) {};
  \ctikzset{resistor = american}
  \end{tikzpicture}
  \caption{ \small RC-network realization
   of impedance $Z(s)\!=\! 1 \! +\! \frac{2}{s+1} \! + \! \frac{1}{s+3}$} \label{fig:Z-Network}
\end{figure}

\begin{figure}[!h]
  \centering
  \ctikzset{bipoles/resistor/height=0.3}
  \ctikzset{bipoles/resistor/width=0.5}
  \begin{tikzpicture} [scale=0.8,transform shape, every node/.style={scale=0.8}]
  \draw (1.5,0) -- (5.5,0);
  \draw (5.5,-2.5) -- (1.5,-2.5); 
  \draw (5.5,0) to[R=$3$] (5.5,-1.) to[L, l=$1$] (5.5,-2.5); 
  \draw (4,0) to[R=$1/2$] (4,-1) to[L, l={$1/2$}] (4,-2.5);
  \draw (2.5,0)  to[R=$1$] (2.5,-2.5);
  \node[] at (0.75,-1.25) {$Y(s)$};
  \draw[->]  (1.25,-1.25) -- (1.5,-1.25);
  \node[circ] at(1.5,0) {}; 
  \node[circ] at(1.5,-2.5) {};
  \ctikzset{resistor = american}
  \end{tikzpicture}
  \caption{\small RL-network realization of
   admittance $Y(s)\! =\! 1 \! +\!  \frac{2}{s+1}\! +\! \frac{1}{s+3}$} \label{fig:Y-Network}
 \vspace*{-2mm}
\end{figure}
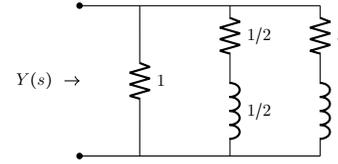

If we choose the transfer function as the impedance $Z(s):=G(s)$ of the realized network, then the system is realized as a RC-network (Foster-I form) as shown in Fig.~\ref{fig:Z-Network}.
If for the RC-network shown in Fig.~\ref{fig:Z-Network} we
choose the states as the voltages across the capacitors suitably scaled $x_i=\sqrt{C_i}v_i$, input as current $I$ injected through the terminals and output as the voltage $V$ across the terminals, then we get a symmetric state-space realization of the transfer function $G(s)$
\begin{equation*}
  \begin{array}{cl}
  A_Z &=\begin{bmatrix} -1 &0 \\0 &-3 \end{bmatrix},\;
  B_Z =\begin{bmatrix}\sqrt{2} \\ 1 \end{bmatrix}=C_Z^T,\;D_Z=1.
  \end{array}
\end{equation*}
When realizing $G(s)=:Y(s)$ as the admittance of a network,
then an RL-realization (Foster-II form) of $G(s)$
is given as Fig.~\ref{fig:Y-Network}.
The impedance of the RL-network in Fig.~\ref{fig:Y-Network} gives us
the inverse transfer function $G(s)^{-1}$:
\[
G(s)^{-1}=\frac{(s+1)(s+3)}{(s+2)(s+5)}=1 - \frac{\frac{1}{3}}{s+2}-\frac{\frac{8}{3}}{s+5}.
\]
For the RL-network shown in Fig.~\ref{fig:Y-Network}, if the states are chosen as
the currents along the inductors suitably scaled $x_k=\sqrt{L_k}i_k$, input as current $I$ injected through the terminals and output as the voltage $V$ across the terminals, then we get a symmetric state-space realization of the inverse system with transfer function $G(s)^{-1}:(A_Y,B_Y,C_Y,D_Y)$:
\begin{equation*}
\begin{array}{cl}
A_Y =\begin{bmatrix} -3 &-\sqrt{2} \\ -\sqrt{2} & -4\end{bmatrix},\;
B_Y=\begin{bmatrix} \sqrt{2} \\ 1 \end{bmatrix},\; C_Y=-B_Y^T,\;D_Y=1.
\end{array}
\end{equation*}
It can be verified that
\[
A_Y\!=\!A_Z-B_ZD_Z^{-1}C_Z, B_Y\!=\!B_ZD_Z^{-1}, C_Y\!=\!-D_Z^{-1}C_Z, D_Y\!=\!D_Z^{-1},
\]
and we pursue this in more generality for MIMO systems later below.

An important problem is that of optimal charging and discharging
i.e. charging the circuit to a specified state with the minimum supply of energy 
from the (multi-)port and
that of discharging the circuit from a specified state with maximum energy
extraction from the (multi-)port.
The energy required for charging and the energy extractable by discharging
are given by the solutions of an appropriate Algebraic Riccati equation (ARE), pursued
later below.
The current/voltage trajectories corresponding optimal charging and discharging
are governed by, respectively, the antistable and stable spectral zeros of the system.
If the spectral zeros are real then the trajectories are purely exponential,
but if two or more of the spectral zeros are nonreal,
then the optimal trajectories would contain oscillations.
In fact, it is easily verified that for RLC systems with two or more
system-poles/zeros on the imaginary axis $j\R$,
some spectral zeros also lie on the imaginary axis $j\R$ and hence the
optimal charging/discharging trajectories are oscillatory.
Hence an important question arises naturlaly for passive systems:
when does a system have only {\em exponential (and non-oscillatory)} optimal
charging/discharging trajectories? 
Note that this is the same as the question: when does a passive system
have only real spectral zeros?

Further, continuing with the property of zeros-interlacing-poles (ZIP) property,
whose study has primarily been restricted to SISO systems,
this paper relates MIMO systems with symmetric state-space realizations and
the ZIP property, using the appropriate notion of system-zero, and also 
relates their interlacing with that of spectral zeros.

\subsection{Contributions of the paper} \label{subsec:summary-contri}
\noindent
In this section, we summarize the contributions in this paper.
In Section~\ref{sec:new_balancing_results}, we study balancing
of strictly passive systems using extremal solutions
of its Algebraic Riccati Equation (ARE)
and propose new notions of positive real quasi-balancing.
In particular,
\begin{itemize}
\item We propose two forms of positive real quasi-balanced realization:
  Form-I ($\Kmax=I$ and $\Kmin$-diagonal), here all the states of length 1
 require equal energy to reach while energy that can be extracted from a state
 is conveyed by diagonal entry of $\Kmin$; and Form-II ($\Kmin=I$ and $\Kmax$-diagonal)
  equal energy can be extracted from each of the states of length 1,
  while the energy required to reach each state is conveyed by diagonal entry of $\Kmax$.
  \item We formulate similarity-transformations for obtaining
  positive real quasi-balanced realizations from a given state-space
  realization and also from one form to another.
  \item We prove the inter-relation between singular values associated
  to  the two forms
  of positive real quasi-balancing and positive real balancing.
  \item We finally prove that a strictly passive system in
  a symmetric state-space realization is
  positive-real balanced: Lemma~\ref{thm:prop:symmetric_prb}.
\end{itemize}

\noindent
In Section~\ref{sec:siso:interlacing} we study 
spectral-zero properties for strictly passive \emph{SISO} systems. 
\begin{itemize}
  \item We first show that for a strictly passive SISO system which
  admits a symmetric state-space  realization, all the spectral-zeros are real
  and further the system-poles, system-zeros and spectral-zeros are 
  interlaced with each spectral-zero lying between a pair of
  system-pole/zero: Theorem~\ref{thm:real:spectral:zeros_siso}.
  \item In Lemma~\ref{lem:sum of squares},  we formulate relations between
   the product and sum of squares of the spectral zeros with the system-poles and system-zeros.
  \item We also show as a special case that for single-order SISO systems,
   the spectral-zero is the geometric mean of the system-pole and system-zero.
\end{itemize} 

\noindent
As mentioned in Section~\ref{subsec:literature}, though SISO systems with
zeros-interlacing-poles (ZIP) property have
been well-studied, extensions have seldom been pursued for MIMO systems;
even recent papers dealing with ZIP property are limited to SISO systems only.
In Section~\ref{sec:mimo:interlacing} we formulate and extend many properties
of spectral zeros for MIMO systems.
In addition to proving the SISO results for the MIMO case (under appropriate
conditions), we also show that
\begin{itemize}
  \item  for symmetric state-space systems, not only are the
   system-poles and system-zeros are interlaced, but the spectral-zeros
   are also interlaced between each pair of
  system-pole/zero: Theorem~\ref{thm:SZ with ZIP MIMO},
  \item a strictly passive MIMO system with a symmetric state-space realization, in which
  the feed-through matrix $D$ can be scaled, exhibits ZIP for sufficiently 
  large $D$: Lemmas~\ref{lem:sufficiently-large-D-for-Z}
  and \ref{lem:sufficiently-large-D-for-Y}.  
\end{itemize} 

\subsection{Organization of the paper}
\noindent
The rest of the paper is organized as follows.
Section~\ref{sec:preliminaries} contains some preliminaries required for the paper. 
In Section~\ref{sec:new_balancing_results} we present and prove some new results 
in ARE-solution based balancing of strictly passive MIMO systems. 
Section~\ref{sec:siso:interlacing} contains the main results for strictly
passive SISO systems:
interlacing properties of system-zeros, system-poles and spectral zeros.
We then extend the interlacing properties to MIMO systems
in Section~\ref{sec:mimo:interlacing}.
Section~\ref{sec:examples} contains some examples
that illustrate the main results of the paper.
Finally, Section~\ref{sec:conclusion} contains concluding remarks.

\section{Preliminaries} \label{sec:preliminaries}
In this paper we consider linear time-invariant dynamical system $\Sigma$ 
with minimal i/s/o representation $(A,B,C,D)$ and transfer function $G(s)$.
\begin{align} \label{eqn:StateSpace}
\hspace*{-1.5mm} \Sigma:\!  \left\{\!\!
 \begin{array}{l}
\dot{x}(t)=Ax(t) + Bu(t)\\
y(t)=Cx(t)+Du(t)
\end{array},\!
\right. 
\;G(s) \! = \! C(sI-A)^{-1}B+D
\end{align}
where $A \in \Rnn,B \in \Rnp,C \in \Rmn,D \in \Rmp$.
In this paper we consider passivity and hence
systems with $m=p$, and thus $D$ is square.
Further, we assume
$B$ is full column rank and $C$ is full row rank: this rules out
redundancy in inputs/outputs. We also assume that $n>m$.

\subsection{Passivity and positive realness}
Passive systems are a class of systems which contain no source
of energy within, but only absorb externally supplied
energy; they however can store energy supplied externally in the past. 
Passive and strictly passive systems defined below.
\begin{definition} 
  A system $ \Sigma $ is said to be passive if 
  \begin{equation*}
  \int_{-\infty}^{t} u(\tau)^T y(\tau)\: \mathrm{d}\tau  \geqslant  0 \quad \text{for all}\; t \in \mathbb{R}\; \text{and all} \;u \in \mathcal{L}_2(\mathbb{R}).
  \end{equation*}
  The system $\Sigma$ is strictly passive if there exists $ \delta > 0 $ such that
  \begin{equation*}
  \int_{-\infty}^{t}\!\!\! u(\tau)^T y(\tau) \mathrm{d}\tau  \! \geqslant \!  \delta \! \! \!
  \int_{-\infty}^{t}\!\!\! u(\tau)^T u(\tau) \mathrm{d}\tau \mbox{ for all }t \in \mathbb{R},u \in  \mathcal{L}_2(\mathbb{R}). 
  \end{equation*}
\end{definition}
\noindent
There are various definitions of strict passivity \cite[Chapter~$6$]{khalil 2002},
the definition we used above has been termed strict input-passivity.
For LTI systems, positive realness of the transfer matrix is linked to passivity.
\begin{definition} \cite{anderson2013network}
  A real rational transfer function matrix $ G(s) $ is said to be positive real 
  if $G(s)$ satisfies:
  \begin{enumerate}
    \item $ G(s) $ is analytic for $ \Re(s) > 0 $,
    \item $ G(s) + G(s)^* \geqslant 0 $ for all $ \Re(s) > 0 $.
  \end{enumerate}
\end{definition} 
\noindent
It is well-known that an LTI system is passive if and only if 
its transfer function matrix is positive real \cite[Lemma~6.4]{khalil 2002}
and, further, for such systems with a state-space realization $(A,B,C,D)$, we have
$(D+D^T) \geqslant 0$.
In addition, for strictly passive systems, none
of the system-poles/zeros lie on the imaginary axis and $(D+D^T)>0$. 

\subsection{Spectral zeros}
The spectral zeros of a positive real system
with transfer function $G(s)$ are defined
as $\MU \in \mathbb{C} $ such that: 
\begin{align*}
&\mbox{det} [G(\MU) + G(-\MU)^T]=0.
\end{align*} 
Considering controllable and observable $n$-th order
systems for which $(D+D^T)$ is invertible, the spectral zeros counted with their multiplicities are exactly the eigenvalues of the  Hamiltonian matrix $H \in \R^{2\n \times 2\n}$ defined as:
\begin{align} \label{eqn:Hamiltonian}
\hspace*{-3mm} H:& \! \! = \begin{bmatrix}
A-B(D+D^T)^{-1}C & B(D+D^T)^{-1  }B^T         \\
-C^T(D+D^T)^{-1}C & -(A-B(D+D^T)^{-1}C)^T        
\end{bmatrix}\!. \!\!
\end{align}
The spectral zeros are symmetric about the imaginary axis $j\R$. Considering a strictly passive system, $H$ does not have any eigenvalues on the imaginary axis $j\R$, and
there are $2\n$ spectral zeros of the system of which $n$-spectral zeros are in the $\C^-$ plane and their $\n$ mirror images in $\C^+$ plane. \\
For example consider a system $\Sigma$ with transfer function $G(s)=\frac{n(s)}{d(s)}=\frac{(s+1)(s+2)}{(s+3)(s+4)}$, the spectral-zeros $\MU \in \C$ satisfy: 
\begin{equation*}
\Scale[1.2]{  \begin{array}{rl}
    \frac{n(s)}{d(s)}+\frac{n(-s)}{d(-s)}=\frac{n(s)d(-s)+n(-s)d(s)}{d(s)d(-s)}\!\!&\!\!=\!0,\\
    \Rightarrow \frac{(s+1)(s+2)(-s+3)(-s+4)+(-s+1)(-s+2)(s+3)(s+4)}{(s+3)(s+4)(-s+3)(-s+4)}\!\!&\!\!=\!0.
  \end{array} }
\end{equation*}
Therefore, the spectral-zeros of the system $\Sigma$ are the roots of $\SZP(s)=n(s)d(-s)+n(-s)d(s)=2s^4-14s^2+48$, i.e $\MU=\{2.05+0.84j,\;2.05-0.84j,\;-2.05+0.84j,\;-2.05-0.84j\}$.\\
The system $\Sigma$ can be represented by the state-space
realization
   $A=\big[\begin{smallmatrix}
           -3 &0\\0 &-4
         \end{smallmatrix} \big],
 B=\big[\begin{smallmatrix}2\\-6 \end{smallmatrix}\big],C=\big[\begin{smallmatrix}1 &1 \end{smallmatrix}\big], D=1$.
The eigenvalues of the Hamiltonian matrix $H$ of the system $\Sigma$ as defined
in Eqn.~\eqref{eqn:Hamiltonian} are exactly same as the
spectral zeros: $\MU=\lambda(H)=\{\pm2.05\pm0.84j\}$.

For a strictly passive system $\Sigma$, of order-$n$, we denote the
complex spectral zeros as $\MU(\Sigma)=(\pm\mu_1,\pm\mu_2,\ldots,\pm\mu_n)$
with $\Re(\mu_i)<0$.
We denote the set of stable spectral zeros by $\MU(\Sigma)^-$ with individual
elements being $\MU_i(\Sigma)^-=\mu_i$ and the set of anti-stable spectral
zeros as $\MU(\Sigma)^+$ with elements $\MU_i(\Sigma)^+$.
This paper focusses on formulating conditions such that 
systems have real spectral zeros.
\subsection{Symmetric state-space realization} 
We define a symmetric state-space
realization \cite{anderson2013network}, \cite{liu1998model} as:  
\begin{definition} \label{def:symmetric state-space realizable}
A state-space realization $(A,B,C,D)$ is said to 
be state-space symmetric if  
\begin{equation} \label{eqn:Symmetric} 
  A = A^T, \quad D=D^T \mbox{ and, either $B = C^T$ or $B=-C^T$.}
\end{equation}
\end{definition}

\noindent
If a system with a given state-space realization can be transformed
into the above form, then we call that system symmetric state-space realizable.
State-space symmetric systems have been called 
{\em internally symmetric} \cite{willems1976realization} and are 
distinct from so-called externally symmetric systems
where $G(s)=G(s)^T$. 
Passive systems which admit symmetric state-space realization
are part of a broader class of systems called 
relaxation systems \cite{willems1976realization}. 
These systems correspond to physical systems which have only one ``type'' of 
energy storage possibility, e.g. only potential energy or only kinetic energy, but not both. 
Another family of examples which have only one type of storage is 
that of RC or RL electrical networks.
It is easily verified that a symmetric state-space realization
helps in showing that the system-poles and system-zeros are real.  
It has also been shown that SISO systems with zeros interlacing poles admit a
symmetric state-space realization \cite{srinivasan1997model}: we pursue this next.

\subsection{SISO Zero-Interlacing-Poles (ZIP) systems}
A strictly passive SISO system $\Sigma$ with a transfer
function $G(s)$ (appropriately scaled to have $D=1$)
having real system-poles $p_i<0$ and system-zeros $z_i<0$
can be written as:
\[
G(s)=\frac{n(s)}{d(s)}=\frac{(s-z_1)(s-z_2)\cdots(s-z_n)}{(s-p_1)(s-p_2)\cdots (s-p_n)}~.
\]
The system $\Sigma$ is said to have zeros-interlacing-poles (ZIP) property if ordered sets of system-poles/zeros follow either
\[
\begin{array}{l}
z_1<p_1<z_2<\cdots<p_{n-1}<z_n<p_n<0\;\;:(z_i<p_i) ~ \mbox{ ~ or }  \\
p_1<z_1<p_2<\cdots<z_{n-1}<p_n<z_n<0\;\;:(p_i<z_i).
\end{array}
\]
It is evident that if a SISO system $\Sigma$ with transfer function $G(s)$ exhibits ZIP property then the inverse system given by the transfer function
$G(s)^{-1}$ also has the ZIP property.
If $G(s)$ follows ZIP with $p_i<z_i$ then $G(s)^{-1}$ follows ZIP with $z_i<p_i$ and vice-versa.
It is known (see for example \cite{srinivasan1997model})
that strictly passive SISO systems having ZIP can be written in the form 
\begin{equation} \label{eqn:pfss_symmetric_system}
G(s)=g_\infty + \sum_{k=1}^{k=n} \frac{g_k}{s-p_1}
\end{equation}
where  $g_\infty >0$, $p_1 <\cdots < p_n<0$, and
\[
g_k > 0\mbox{ if } z_i<z
p_i, \mbox{ and }
g_k < 0 \mbox{ if } p_i<z_i~.
\]
Further, such systems admit a symmetric
state-space realization~\cite{willems1976realization} given as
\begin{equation}\label{eqn:symmetric state-space}
\begin{array}{cl}
A &=\diag(p_1,p_2,\ldots,p_n),\\ 
B^T &=[|g_1|^{\frac{1}{2}} \;|g_2|^{\frac{1}{2}}\;\cdots \;|g_n|^{\frac{1}{2}}],\; C=\pm B^T, \;  D=g_\infty
\end{array}
\end{equation}
with $B=C^T$ if $g_k>0$, and $B=-C^T$ if $g_k<0$.\\

Symmetric state-space systems have been well-studied in the literature.
A class of well-studied systems with 
collocated actuators and sensors \cite{sun1996,DoJefDan1992,HalDunRez2003} 
result in $B=C^T$. 
Collocated sensors and actuators in decentralized control systems reduce the 
complexity and hence are economically advantageous. Symmetry within $A$ arises
due to, for example, a certain type of reciprocity in the interaction between
subsystems in a network of such simpler systems:
multi-agent networks with single integrator have been modelled to
obtain a symmetric state-space realization \cite{briegel_2011},\cite{yoon_2013}. 

\subsection{Algebraic Riccati equation}
The algebraic Riccati equation (ARE) for a system $\Sigma$ in minimal
i/s/o realization $(A,B,C,D)$ with respect to the passivity supply rate $u^T y$ is
\begin{equation} \label{eqn:ARE}
A^TK + KA + (KB-C^T)(D+D^T)^{-1}(B^TK-C)\! =\! 0.
\end{equation} 
By the well-known KYP lemma, the system $\Sigma$ is positive real
if and only if there 
exists a positive definite solution $K=K^T$ to the above equation. 
The set of ARE solutions is known to be 
a bounded and finite set with a maximum $\Kmax $
and a minimum $\Kmin$:
$0 \;<\;\Kmin\;\leqslant \;K\;\leqslant \;\Kmax$.
The solutions of the ARE in Eqn.~\eqref{eqn:ARE} can be computed from an
$n-$dimensional invariant subspace $\subset \R^{2n}$
of the associated Hamiltonian matrix, $H$ as follows
\begin{align} \label{eqn:AREsoln}
H
\left[\begin{array}{c}
X\\
Y\\
\end{array}\right] = 
\left[\begin{array}{c}
X\\
Y\\
\end{array}\right]R  \mbox{ ~ and define ~ } K:=YX^{-1}
\end{align}
where $X,Y \in \Rnn$,  $R \in \Rnn$ (for the real eigenvalue case)
is an upper triangular matrix with diagonal
as $n$ eigenvalues of the Hamiltonian matrix, i.e. $n$-spectral zeros. 
Each solution $K$ can be associated with $n$-spectral zeros chosen from $2n$ spectral zeros. 
When either $n$ stable or $n$ anti-stable spectral zeros are chosen, 
we get the ARE's {\em extremal solutions}:
\begin{align} \label{eqn:H-invariantsubspace}
\hspace*{-3mm} H
\left[\begin{array}{c}
X_+\\
Y_+
\end{array}\right] = 
\left[\begin{array}{c}
X_+\\
Y_+\\
\end{array}\right]R_+  \text{ and }
H
\left[\begin{array}{c}
X_-\\
Y_-
\end{array}\right] = 
\left[\begin{array}{c}
X_-\\
Y_-\\
\end{array}\right]R_- 
\end{align}
where $X_{\displaystyle \pm},Y_{\displaystyle \pm} \in \Rnn$ with
$ \Re(\lambda(R_+))>0$ and $ \Re(\lambda(R_-))<0 $.
Then, $\Kmax=Y_+X_+^{-1}$ and $\Kmin=Y_-X_-^{-1}$.

\subsection{Ordering convention} \label{subsec:ordering convention}
We frequently require comparison between elements of multiple sets of real numbers (like eigenvalues of symmetric
matrices), and it helps to have an ordering and indexing convention for such sets. Suppose $X$ is the set of 
eigenvalues of an $n\times n$ real symmetric matrix, i.e. elements of $X$ are real, 
and with possible repetitions. Order and index the
elements $\lambda_1$, $\lambda_2$, $\ldots$, to satisfy
\begin{equation} \label{eqn: eigenvalues ordering}
\lambda_{\min} = \lambda_1\;\leqslant \; \lambda_2\;  \leqslant\; \ldots \;\leqslant \lambda_n = \lambda_{\max}.
\end{equation}
In this context, we also need the $n-1$ successive differences, which
we denote by $\nu_i$, i.e. 
\[
\nu_i = \lambda_{i+1}-\lambda_i \mbox{ for } i=1,\ldots,n-1
\]
with $\nu_{\min}$ and $\nu_{\max}$ being the minimum and
maximum of these $(n-1)$ successive differences. 

\section{New methods in ARE-solution based balancing} \label{sec:new_balancing_results}

\noindent
In model order reduction studies, a widely used tool is the notion
of balancing of a system.
In this section we focus on balancing of system with respect to
the extremal solutions of the ARE: $\Kmax$ and $\Kmin$.
State-space realizations which are balanced with respect
to such energy functions reveal the energy-wise significance of the states. 
The extremal positive definite solutions of the ARE $\Kmin$ and $\Kmax$ have special significance in terms of the energy dissipation by the system. 
For a state-value $a \in \Rn$, consider $\B_a$, the set of all
continuous system trajectories $(u, x, y)$ which are zero outside
a finite interval satisfying equation \eqref{eqn:StateSpace} and with $x(0) = a$. 
Then,
\begin{align*}
a^T \Kmax a\;=\; \inf_{\substack{(u,x,y) \in \B_a, \\ x(-\infty)=0}}\quad \int_{-\infty}^{0}2uy \;dt, \\
a^T \Kmin a\;=\; \sup_{\substack{(u,x,y) \in \B_a, \\ x(\infty)=0}}\quad \int_{0}^{\infty}-2uy \;dt.
\end{align*}
Thus $a^T \Kmax a $ is the minimum energy required to 
reach a state $x(0)=a$ from the state of rest $x(-\infty)=0$
while $ a^T \Kmin  a$ is 
the maximum energy that can be extracted as the system is brought to rest $x(\infty)=0$
from state $x(0)=a$.
 Positive Real Balancing of passive systems has been a popular tool for passivity preserving model reduction \cite{antoulas2001survey}. We first present some new results of positive real balancing in systems with symmetric state-space realization, then we introduce positive real quasi-balancing. 
\begin{definition}\label{def:Positive Real Bal}
  \cite[Section 7.5.4]{antoulas2001approximation}
  A positive real MIMO system $\Sigma$ with i/s/o representation $(A,B,C,D)$ is 
  said to be in positive real balanced realization if the extremal solutions 
  of the ARE, $\Kmax$ and $\Kmin$, are 
  related as 
  \begin{equation*} 
  \Kmax = \Kmin^{-1}~.
  \end{equation*}
   If $\Kmax$ and $\Kmin$ are
 simultaneously diagonalized\footnote{\label{ft:simulataneously}
     Since both $\Kmax$
    and $\Kmin$ are symmetric and positive definite, they
    can be {\em simultaneously diagonalized} by a congruence
    transformation, i.e. there exists a
    suitable basis in which the quadratic forms
    corresponding to matrices $\Kmax$ and $\Kmin$
    are both diagonal.}
then:
  \begin{equation*}
  \Kmin\!=\!\Kmax^{-1}\!=\!\diag(\sigma_1, \sigma_2,\ldots,\sigma_n)\;\! \mbox{ with }0 < \sigma_1 \leqslant \cdots \leqslant \sigma_n \leqslant 1.
  \end{equation*} 
   The $\sigma_i$ are called the positive real singular values of $\Sigma$.
\end{definition}

We present a new form of balancing in positive real systems with respect to the extremal storage functions: $\Kmax$ and $\Kmin$.  If a system is balanced with respect to $\Kmax$ then
the amount of energy required to reach any state (of unit-length $\|a\|_2 = 1$)
is the same i.e.  $\Kmax=I$ and $\Kmin$ is a diagonal matrix. 
Similarly, if a positive real system is balanced with respect to $\Kmin$ then
the amount of energy that can be extracted from any
state (again of unit-length)
is the same i.e. $\Kmin=I$ and $\Kmax$ is diagonal. 
We call this positive real \underline{quasi-balancing} and there are two
forms of this balancing when $\Kmax=I$ or when $\Kmin=I$.

\begin{definition} \label{lem:quasi-balanced}
  A positive real MIMO system $\Sigma$ is said to be in positive real quasi-balanced form 
  if one of the extremal positive definite solutions of the ARE is identity. 
  Positive real quasi-balanced Form-I if $\Kmax =I$ and 
  positive real quasi-balanced Form-II if $ \Kmin =I$.
\end{definition}

Our first main result of this section states that one can always obtain 
a positive real state space system in these forms.

\begin{theo} \label{thm:balancings1}
  A strictly passive MIMO system $\Sigma$, admits a Form-I positive real quasi-balanced 
  realization $(A^{+},B^+,C^+,D^+)$ such that $\Kmax^+=I$ and 
  $\Kmin^+=\Lambda^+ = \diag (\sigma^+_1, \ldots,  \sigma^+_n)$ where $0<\sigma^+_1 \leqslant \sigma^+_2, \leqslant \cdots \leqslant \sigma^+_n \leqslant 1$. 
\end{theo}

\begin{proof}
  Consider a strictly passive system $\Sigma:(A,B,C,D)$ with extremal storage functions $\Kmax$ and $\Kmin$. Since both $\Kmax$ and $\Kmin$ are symmetric and
positive definite, they can be simultaneously diagonalized
(see Footnote~\ref{ft:simulataneously}).
Compute the Cholesky factorization of
 $\Kmax$, i.e. $\Kmax=:R^T R$ and choose $ S:=R^{-1}$.
Next compute: $P:=S^T \Kmin S $.
Since $ P$  is symmetric, we write: $ P=:Q \Lambda^+ Q^T $, with $Q$-orthogonal.
We next define the transform matrix: $T:=SQ $.
The system $ \Sigma:(A,B,C,D) $ with a transformed state $  x^+=Tx $ is given as:
$A^+:=T^{-1}AT; \;\; B^+:=T^{-1}B;\;\; C^+:=CT;\;\; D^+:=D $.

Now in this basis transform the extremal storage functions are given as:
$\Kmax^+=T^T \Kmax T =(SQ)^T \Kmax (SQ)=Q^T (R^{-1})^T (R^T R) (R^{-1} Q)=Q^T Q $
which implies $\Kmax^+=I $;
$\Kmin^+=T^T \Kmin T =(SQ)^T \Kmin (SQ) =Q^T (R^{-1})^T \Kmin (R^{-1} Q) =Q^T P Q $
$\implies \Kmin^+= \Lambda^+ =\diag(\sigma^+_1,\;\cdots\;,\sigma^+_n)$.
\end{proof}

\noindent
The $\sigma_i^+$ are called the Form-I positive real quasi-singular values and
if they are distinct then the positive real quasi-balanced realization $(A^{+},B^+,C^+,D^+)$
can be shown to be unique.
\begin{theo} \label{thm:balancings2}
  A strictly passive MIMO system $\Sigma$, admits a Form-II positive real quasi-balanced 
  realization $(A^{-},B^-,C^-,D^-)$ such that $\Kmin^-=I$ and $\Kmax^-=\Lambda^-=\diag(\sigma^-_1,\cdots, \sigma^-_n)$ where $1 \leqslant \sigma^-_1 \leqslant \sigma^-_2, \leqslant \cdots \leqslant \sigma^-_n$.
\end{theo}
\noindent
The proof of Theorem~\ref{thm:balancings2} is analogous to the earlier proof, hence omitted.
The $\sigma_i^-$ are called the Form-II positive real quasi-singular values.
The positive real quasi-singular values $\sigma_i^-$ and $\sigma_i^+$ are related
with the positive real singular values $\sigma_i$ by the following lemma.
\begin{lemma}\label{lem:singular values relation}
For a strictly passive MIMO system  $\Sigma:(A,B,C,D)$ the positive real singular values are related with the positive real quasi-singular values $\sigma^+_i$ and $\sigma^-_i$ as:
\begin{equation*}
\sigma_i=\; \sqrt{\sigma^-_i}\; =\; \frac{1}{\sqrt{\sigma^+_i}}.
\end{equation*}
\end{lemma}
\noindent
The proof of Lemma~\ref{lem:singular values relation} is straightforward and hence
omitted.

\begin{theo} \label{thm:FormI2II}
  A strictly passive MIMO system $\Sigma$ in Form-I positive real quasi-balanced realization $(A^{+},B^+,C^+,D^+)$ can be transformed to Form-II quasi-balanced realization $(A^{-},B^-,C^-,D^-)$ by the transformation matrix:
  \begin{equation*}
  T=(\Lambda^+)^{-\frac{1}{2}}=\diag(\frac{1}{\sqrt{\sigma^+_1}},\;\frac{1}{\sqrt{\sigma^+_2}},\;\ldots,\;\frac{1}{\sqrt{\sigma^+_n}})~. 
  \end{equation*}
  Further, $(A^{-},B^-,C^-,D^-)$ are given by
  \begin{equation*}
  A^-:=T^{-1}A^+T; \;\; B^-:=T^{-1}B^+;\;\; C^-:=C^+T;\;\; D^-:=D^+.
  \end{equation*}
\end{theo}

\noindent
Before we proceed with the proof, we note that, analogous to the above result,
a strictly passive system $\Sigma$ in Form-II positive real
quasi-balanced realization $(A^{-},B^-,C^-,D^-)$ can be transformed
to Form-I quasi-balanced realization $(A^{+},B^+,C^+,D^+)$ by the
similarity transformation matrix:
$\diag(\frac{1}{\sqrt{\sigma^-_1}}, 
\frac{1}{\sqrt{\sigma^-_2}},\dots, \frac{1}{\sqrt{\sigma^-_n}})$; we do not 
prove this part due to the close parallel to the proof below (of Theorem~\ref{thm:FormI2II}).

\begin{proof}
   Consider a strictly passive MIMO system $\Sigma$ in Form-I positive real  quasi-balanced realization $(A^{+},B^+,C^+,D^+)$ then,
\[
   \Kmax^+=I,\; \Kmin^+=\Lambda^+=\diag(\sigma^+_1,\;\sigma^+_2,\;\ldots,\;\sigma^+_n)~.
\]
   As $\Kmax$ and $\Kmin$ are quadratic forms, a basis transformation of the state-space is congruence transform for them. If we choose the basis transform matrix as:
\[
    T=(\Lambda^+)^{-\frac{1}{2}}= \diag(\frac{1}{\sqrt{\sigma^+_1}},\;\frac{1}{\sqrt{\sigma^+_2}},\;\ldots,\;\frac{1}{\sqrt{\sigma^+_n}})~. 
\] 
then the congruence transform of the $\Kmax$ and $\Kmin$ using $T$ results in:
\begin{equation*}
\begin{array}{cl}
   T^T \Kmin^+ T &=(\Lambda^+)^\frac{-1}{2} \Lambda^+ (\Lambda^+)^\frac{-1}{2} = I =:\Kmin^-,\\
    T^T \Kmax^+ T &=(\Lambda^+)^\frac{-1}{2} I (\Lambda^+)^\frac{-1}{2}=(\Lambda^+)^{-1}=:\Kmax^-.
\end{array}
\end{equation*}
Therefore, the system in Form-II positive real quasi-balanced realization is given by
$ A^-:=T^{-1}A^+T, \; B^-:=T^{-1}B^+, C^-\! :=\! C^+T$ and $D^- := D^+.$
\end{proof}
\begin{corollary}
  For a strictly passive MIMO system $\Sigma$ the extremal storage functions, in the positive real quasi-balanced realizations Form-I and Form-II are related as:
\[
  \Kmax^+ \;=\;\Kmin^-\;=I, \mbox{ and } \Kmax^- = (\Kmin^+)^{-1}.
\]
\end{corollary}

\noindent
The next result regarding symmetric state-space realizations follows by
using Definition~\ref{def:Positive Real Bal} of positive-real balancing,
and by straightforward verification of balancing.
\begin{theo} \label{thm:prop:symmetric_prb}
  A strictly passive MIMO system having
a symmetric state-space realization is positive real balanced.
\end{theo}
\begin{proof}
  Consider first a strictly passive system $\Sigma$ in state-space 
  symmetric realization with $B= + C^T$ (and $A=A^T,D=D^T$).
  Let $K=K^T$ be a positive definite solution of 
  the ARE, then pre-multiplying and post-multiplying the ARE Eqn.~\eqref{eqn:ARE} 
  by $K^{-1}$ we get:
  \begin{equation*}
  K^{-1}A^T \!+\! AK^{-1}+(B-K^{-1}C^T) (D+D^T)^{-1}(B^T-CK^{-1})\! =\!0.
  \end{equation*}
  and after rearranging the matrices and using Eqn.~\eqref{eqn:Symmetric} we get
  \begin{equation*} \label{eqn:K inverse}
    A^TK^{-1}\! +\! K^{-1}A \!+\! (K^{-1}B-C^T) (D+D^T)^{-1}(B^TK^{-1}-C)\! =\! 0. 
  \end{equation*}
  Therefore, if $K$ is a solution of the ARE then $K^{-1}$ is also a solution. If $\Kmax$ is the maximal solution then it implies that $\Kmax^{-1}$ is the minimal solution. It follows that
\begin{equation*}
  \Kmax= \Kmin^{-1}.
\end{equation*}
Similarly, it can be verified along the same lines that
if the given symmetric state-space realization satisfies $B=- C^T$, then
too, both $K$ and $K^{-1}$ satisfy the ARE.
This completes the proof of the theorem. 
\end{proof}

\section{Interlacing properties in SISO systems' spectral zeros} \label{sec:siso:interlacing} 

In this section, we focus on SISO systems since the proof techniques 
are simpler and offer more insight. Many of these results are extended
under appropriate assumptions to the MIMO case in the following section: those
results use different proof-techniques, namely,
those involving interlacing properties between eigenvalues of
pairs of symmetric matrices.
In this section, we first formulate a result about passive SISO systems which have
only real spectral zeros, one of the main results of this section,
Theorem~\ref{thm:real:spectral:zeros_siso}.
The following lemma is helpful for proving this main result.

\begin{lemma}\label{lem:no:complex}
  Consider the function $f(x): \C \rightarrow \C$ defined by
  \begin{equation}
  f(x):=\sum_{k=1}^{n} \frac{q_k} {x^2-p_k^2}
  \end{equation}
  with $p_k, q_k$ real and $q_k > 0$ for $k=1,\dots,n$.
  Then, $f(x)$ has only real zeros. 
\end{lemma}

\begin{proof} 
  We prove the fact that all the zeros are real by contradiction.
  Suppose a zero $x_1$ of $f(x)$ is written as
  $x_1=a+bj$ with $a,b \in \R$. Evaluating $f(x_1)=0$, we get
  \begin{equation} \label{eqn:lem:lno:complex1}
  \Scale[1.1]{ \frac{q_1}{(a+bj)^2-p_1^2}+\frac{q_2}{(a+bj)^2-p_2^2}+\cdots+\frac{q_n}{(a+bj)^2-p_n^2}=0.}
  \end{equation}
  Now, $(a+bj)^2-p_k^2=(a^2-b^2-p_k^2)+2abj=:u_k+vj$ (say), with $u_k$ and $v$ real.
Therefore the above equation 
can be rewritten as
\begin{equation*} 
  \frac{q_1}{u_1+vj}+\frac{q_2}{u_2+vj}+\cdots+\frac{q_n}{u_n+vj}=0.
\end{equation*}
Simplifying each term of the above equation
by making the denominator real, we get
  \begin{equation*} 
  \frac{q_1(u_1-vj)}{u_1^2+v^2}+\frac{q_2(u_2-vj)}{u_2^2+v^2}+\cdots+\frac{q_n(u_n-vj)}{u_n^2+v^2}=0.
  \end{equation*}
  Since $q_k>0$ and $(u_k^2+v^2)>0$, the imaginary parts of each term in
the above equation 
have the same sign (dictated by $v$) and hence cannot cancel out.
Therefore, the above equation 
is satisfied if and only if $v=0$, equivalently, $ab=0$. 
  If $a=0$ and $b\neq 0$, then it is easily 
  seen that Eqn.~\eqref{eqn:lem:lno:complex1} is not satisfied 
  since each fractions would be real
  and positive (as $q_k<0$). 
  Therefore, if $x_1=a+bj$ is a zero of $f(x)$, then $b=0$. 
Thus $f(x)$ has only real zeros.
\end{proof}

Using the above lemma, we prove the following result that the spectral
zeros of ZIP systems are real, and the spectral-zeros too satisfy an
interlacing property.

\begin{theo} \label{thm:real:spectral:zeros_siso}
Suppose a strictly passive SISO system exhibits the ZIP property.
Then all the spectral zeros are real.\\
Further, assume the sets of system-poles $p_i<0$, system-zeros $z_i<0$ and stable
  spectral zeros $\MU_i<0$ are indexed such that:
  \begin{equation}
  \begin{array}{l}
  p_1 < p_2 < \cdots < p_n<0, \quad
  z_1 < z_2 < \cdots < z_n<0, \\
  \MU_1 < \MU_2 < \cdots < \MU_n<0 
  \end{array}
  \end{equation}
  and assume, without loss of generality,  $z_1 \! < \! p_1$.
  Then, in fact,
  \begin{equation}
  z_1 < \MU_1 < p_1 < z_2 < \MU_2 < p_2<\cdots <z_n<\MU_n < p_n<0~.
  \end{equation}
  In other words, not just are the poles and zeros interlaced, but between
  every such pair of pole-zero, there is also a stable spectral zero.
\end{theo}

\begin{proof}
Due to the assumptions in the theorem,
the transfer function $G(s)$ of the strictly passive SISO system $\Sigma$ can
be represented by Eqn.~\eqref{eqn:pfss_symmetric_system} and
the system-poles $p_i<0$ and system-zeros $z_i<0$ are real, distinct and satisfy:
\begin{equation} \label{eqn:poles_zeros_ordered}
  z_1 < p_1 < z_2 <p_2 < \cdots < p_{n-1} < z_n < p_n~.
\end{equation}
We first prove that all the spectral zeros are real, and then we prove
their interlacing property with system poles and zeros.
Expand in partial fractions the transfer function $G(s)$, and
the spectral zeros of the system are the zeros of $ G(s)+G(-s) $:
  \begin{equation*} 
\Scale[1.1]{
  \begin{array}{rc}
  g_\infty +\frac{g_1}{s-p_1}+\cdots+\frac{g_n}{s-p_n}+g_\infty +\frac{g_1}{-s-p_1}+\cdots+\frac{g_n}{-s-p_n}&=0 \\
  \implies 2( g_\infty +\frac{g_1p_1}{s^2-p_1^2}+\frac{g_2p_2}{s^2-p_2^2}+\cdots+\frac{g_np_n}{s^2-p_n^2})&=0 
  \end{array}}
  \end{equation*}
  where $g_\infty, g_i>0$ and $p_i<0$. Without loss of generality, we assume
$g_\infty =1$.
  Therefore, the above equation can be 
 rewritten as $1+f(s)=0$ and as $g_kp_k <0$ from Lemma~\ref{lem:no:complex}, it has only real zeros. This proves that all spectral zeros are real.
  
Next, write $G(s)+G(-s)=0$ in terms of the system poles and zeros as:
  {\small \begin{equation} \label{eqn:proof_pole_zeros}
    \begin{array}{l}
      \frac{\displaystyle \prod_{i=1}^{n}(s-z_i) \prod_{i=1}^{n}(-s-p_i)+\prod_{i=1}^{n}(-s-z_i) \prod_{i=1}^{n}(s-p_i)}{\displaystyle \prod_{i=1}^{n}(s-p_i) \prod_{i=1}^{n}(-s-p_i) } =0~. 
    \end{array}
  \end{equation} }
  The spectral zeros are the roots of the numerator of the Eqn.~\eqref{eqn:proof_pole_zeros} and therefore it can be seen that there are $2n$ spectral zeros.  Since the spectral zeros are symmetric about the imaginary axis, 
  there are $n$ stable spectral zeros $(\MU_1,\MU_2,\ldots,\MU_n)$ in $\R_-$ and $n$ anti-stable spectral zeros $(-\MU_1,-\MU_2,\dots,-\MU_n)$ in $\R_+$. 
  We consider stable spectral zeros in $\R_-$.
Now, for $\MU_i<0$, the terms $(-s-p_i)$ and $(-s-z_i)$ 
  are positive and their product in Eqn.~\eqref{eqn:proof_pole_zeros} can be replaced
by positive definite functions $r(s)>0$ and $t(s)>0$ for real $s$ and $s<0$.
Therefore the spectral zeros are the roots of polynomial $\SZP(s)$:
  \begin{equation*}
  \begin{array}{cl}
  \SZP(s)&=t(s) (s-p_1)\cdots(s-p_n)+r(s) (s-z_1)\cdots(s-z_n) \\
  \SZP(s)&=:t(s)P_1(s)+r(s)P_2(s)\; \text{       (say).}
  \end{array}
  \end{equation*}
  We next use Bolzano's theorem\footnote{\textbf{Bolzano's theorem:} Suppose the function $f:\R\rightarrow\R$ is continuous in the
interval $(a, b)$ and suppose $f(a)\cdot f(b)<0$. Then there exists an $x_0$ in the
open interval $(a,b)$ such that $f(x_0)=0$.\\
Conversely, if $f(x_1)\cdot f(x_2) > 0 $ for each
$x_1,x_2$  in the interval $[a,b]$, then $f(x)$ has no roots in the interval $[a,b]$.
We say $f$ `does not change sign' in $[a,b]$.} to locate the roots of the
polynomial $\SZP(s)$.

Notice that $\SZP(s)$ is a continuous function in $\R_-$ and the system-poles $p_i$ and system-zeros $z_i$ are indexed as Eqn.~\eqref{eqn:poles_zeros_ordered}. 
  If we consider $s$ in the interval $[p_n, 0]$, $P_1(s) \geqslant 0$
  and $P_2(s)>0$ and hence $\SZP(s)>0$.
Since $\SZP(s)$ does not change sign when $s \in [p_n, 0]$,
there are no roots of $\SZP(s)$ in this interval. 
Similarly, for $s \in (-\infty, z_1]$, the polynomial $\SZP(s)$ does not 
change sign and hence no roots exist in this interval.
When the system-order $n$ is even, for $s \in (p_1,z_2)$, we have $\SZP(s)<0$,
while when $n$ is odd, $\SZP(s)>0$.
As $\SZP(s)$ does not change sign, therefore no roots of $\SZP(s)$ exists
in the interval $[p_1,z_2]$.
Similarly, it can be easily seen that no roots of $\SZP(s)$ exist in
any of the intervals $[p_i,z_{i+1}]$.  
For $s \in [z_1,p_1]$, sign$(\SZP(z_1))=(-1)^{n}$ and sign$(\SZP(p_1))=(-1)^{n-1}$,
i.e. opposing signs, and hence there exists a $\MU_1$ in
the interval $[z_1,p_1]$ satisfying $\SZP(\MU_1)=0$. 
Similarly, it can be shown that in each of the intervals $[z_i,p_i]$,
there exists a $\MU_i$ such that $\SZP(\MU_i)=0$ because there 
  is a sign change in the interval with $\sign(\SZP(z_i))=(-1)^{n-i}$
and $\sign(\SZP(p_i))=(-1)^{n-i+1}$.
  Since there are $n$ intervals
$[p_k,z_k]$, and $n$ spectral zeros
(roots of $\SZP$) in $\R_-$ and each interval has at least one spectral zero,
we conclude that there is exactly one spectral zero in each interval $[p_i,z_i]$.
This proves the required:\\
  $ z_1 < \MU_1 < p_1 < z_2 < \MU_2 <p_2< \cdots <z_n<\MU_n < p_n. $
\end{proof}

The next result relates the spectral zeros with the system poles/zeros.
  \begin{lemma}\label{lem:sum of squares}
    Consider a SISO system $\Sigma$ with biproper transfer function $G(s)$:
    \begin{equation} 
    G(s)=\frac{p(s)}{d(s)}=\frac{(s-z_1)(s-z_2)\cdots (s-z_n)}{(s-p_1)(s-p_2)\cdots (s-p_n)},
    \end{equation}
with all poles and zeros real and negative. Also assume that the poles and
zeros are interlaced. Then, the following hold.
\begin{enumerate}
 \item The product of the $n$-stable/antistable -spectral zeros equals
 the square root of the product of system-zeros and system-poles. Ignoring the
signs, 
\[
| \MU_1\MU_2\cdots \MU_n | = \sqrt{p_1p_2\cdots p_n \cdot z_1z_2\cdots z_n}. 
\]
  \item The sum of the squares of the $n$-stable/antistable spectral-zeros 
$\MU_1^2 + \MU_2^2+\cdots+\MU_n^2$
is 
\[
\sum\limits_{i=1}^n p_i \sum\limits_{i=1}^n z_i
   -  \sum\limits_{i=1}^n \sum \limits_{k=i+1}^n p_i  p_k - \sum\limits_{i=1}^n  \sum \limits_{k=i+1}^n z_i z_k.
\]

 \end{enumerate}
\end{lemma}

\noindent
It may be noted that the claims hold under milder assumptions than
assumed in the above theorem, namely, poles and zeros need not be interlaced, and,
in fact, need not even be real, nor do the spectral zeros have to be real; the
same proof techniques work for the more general case also.
However, since this paper focusses on interlacing properties of poles and zeros
and about real spectral zeros, we do not digress into the general case. We proceed
with the proof of the above result.

\begin{proof}
    
The spectral-zeros are the roots of polynomial
$\SZP(s)$, which is the numerator of $G(s)+G(-s)$, defined by:
\[
    \SZP(s):=\prod_{i=1}^{n}(s-z_i) \prod_{i=1}^{n}(-s-p_i)+\prod_{i=1}^{n}(-s-z_i) \prod_{i=1}^{n}(s-p_i)~.
    \]
Expanding $\SZP(s)$, and noting that only terms with even powers of $s$ remain,
express $\SZP(s)$ as
\begin{equation}\label{eqn:SZP even polynomial}
 \SZP(s)=a_{2n}s^{2n}+a_{2n-2}s^{2n-2}+\cdots+a_2s^2+a_0~.
\end{equation}
    From Theorem~\ref{thm:real:spectral:zeros_siso}, we get that the system has only real spectral zeros. Further, from Eqn.~\eqref{eqn:SZP even polynomial} we get that the spectral zeros occur in pairs and we represent the set as $\{\pm\MU_1,\pm\MU_2,\ldots,\pm\MU_n\}$, with $\MU_i<0$. 
The coefficients of the Eqn.~\eqref{eqn:SZP even polynomial} are given as:
    {\small \begin{equation*}
    \begin{array}{cl}  
    a_{2n}&=2(-1)^n,  \\
    a_{2n-2}&={\displaystyle 
      2(-1)^n( \sum\limits_{i=1}^n \sum \limits_{k=i+1}^n p_i p_k + \sum\limits_{i=1}^n  \sum \limits_{k=i+1}^n z_i z_k- \sum\limits_{i=1}^n p_i \sum\limits_{i=1}^n z_i} ),\\
    \vdots\\
    a_0&={\displaystyle 2\prod_{i=1}^{n}p_i\prod_{i=1}^{n}z_i }~.
    \end{array}
    \end{equation*} }
    Now applying Vieta's Formula\footnote{\textbf{Vieta's Formula:} For a polynomial of degree $n$: $P(x)=a_nx^n+a_{n-1}x^{n-1}+\cdots+a_2x^2+a_1x+a_0$, the sum of the roots of $p(x)$ is equal to $-\frac{a_{n-1}}{a_n}$ and product of the roots is equal to $(-1)^n\frac{a_0}{a_n}$.} we verify that the sum of the spectral-zeros is $0$ as $a_{n-1}=0$, i.e the spectral zeros are symmetrical along the imaginary axis $j\R$. The product of the spectral-zeros is expressed by the coefficients of the polynomial $\SZP(s)$ as:
   { \small \[
    (\MU_1 \MU_2 \cdots \MU_n)(-\MU_1 -\MU_2 \cdots -\MU_n)=(-1)^n\frac{a_0}{a_{2n}}=\prod_{i=1}^{n}p_i\prod_{i=1}^{n}z_i~.
    \] }
   Therefore, we get:
    \begin{equation*}
    | \MU_1\MU_2\cdots \MU_n | = \sqrt{p_1p_2\cdots p_n \cdot z_1z_2\cdots z_n}~.
    \end{equation*}
    Further, if we replace $x=s^2$ in the Eqn.~\eqref{eqn:SZP even polynomial}, we get:
    \begin{equation*}\label{eqn:squared SZP}
    \SZP(x)=a_{2n}x^{n}+a_{2n-2}x^{n-1}+\cdots+a_2x+a_0~.
    \end{equation*}
    There are $n$-roots of $\SZP(x):\{x_1,x_2\ldots,x_n\}$ where $x_i=\MU_i^2$, again applying Vieta's Formula, we get that the sum of the square of the spectral-zeros is:
    {\small \begin{align*}
    &x_i^2+x_2^2+\cdots+x_n^2 = -\frac{a_{n-2}}{a_{2n}},\\
    \implies &\MU_1^2 + \MU_2^2+\cdots+\MU_n^2= \sum\limits_{i=1}^n p_i \sum\limits_{i=1}^n z_i-\sum\limits_{i=1}^n \sum \limits_{k=i+1}^n p_i  p_k - \sum\limits_{i=1}^n  \sum \limits_{k=i+1}^n z_i z_k ~.
    \end{align*} }
This proves  Lemma~\ref{lem:sum of squares}.
\end{proof}

  
\noindent
A special case of the above lemma
is when a SISO system has just one spectral zero: namely
passive SISO systems with only one pole and one zero, the spectral zero is the
geometric mean of the pole and zero values.
\begin{corollary}
Consider a SISO system with transfer function $G(s)=\frac{s-z}{s-p}$, with 
$p,z < 0$, i.e. with only a pair of system-pole/zero. 
Then the stable and anti-stable spectral-zero of the system satisfy
\[
 \pm \MU=\pm \sqrt{pz}.
\]
\end{corollary}

\section{Interlacing properties in spectral zeros
of MIMO systems} \label{sec:mimo:interlacing} 

In this section we pursue MIMO systems and extend several of the results of
the previous section. A first point to note is that for MIMO systems, unlike
the notion of system pole, there are various notions of a system-zero.
While there are some inter-relations (like set-inclusions) between these
various nonequivalent definitions of zeros of a system
\cite{WymSai83}, a natural question is
which notion of zero would possibly yield pole/zero interlacing type
of properties.

In this paper, since we deal with passivity based studies, 
we consider systems with equal number of inputs and outputs.
Hence we assume that the MIMO transfer matrix $G(s)$ is
square and invertible. For such a $G(s)$, we define the system-zeros as
the poles of the transfer matrix $G(s)^{-1}$. Further, we restrict ourselves
to systems in which $G(s)$ is biproper, i.e. the feed-through matrix $D$ in
any state-space realization of $G(s)$ is invertible. Under this
assumption, the state-space equations: $ \dot{x} = Ax + Bu$ and $y=Cx+Du$
can be rewritten as:
\begin{equation} \label{eq:state-space:inverseG}
\begin{array}{rl}
\dot{x} &= (A-BD^{-1}C)x+BD^{-1}y,\\
u&=-D^{-1}Cx + D^{-1}y.
\end{array} 
\end{equation}
The rest of this paper frequently involves dealing with the state-space
representations of $G(s)$ and $G(s)^{-1}$, and to ease notation,
we consider $G(s)$ as say the impedance matrix of a system, say $Z(s)$, and
denote the state-space realization by $(A_Z, B_Z, C_Z, D_Z)$ and hence
the state-space realization of $G(s)^{-1}$, the corresponding admittance
matrix $Y(s)$ as 
in Eqn.~\eqref{eq:state-space:inverseG} by $(A_Y, B_Y, C_Y, D_Y)$. 
For easy reference, we include this as a definition.
\begin{definition} \label{defn:inverseG:state-space}
Consider a MIMO system $\Sigma_Z$ with a biproper transfer matrix $G(s)=Z(s)$
and having a state-space realization $(A_Z,B_Z,C_Z,D_Z)$. 
The
inverse system $\Sigma_Y:(A_Y,B_Y,C_Y,D_Y)$ is defined as:
\[
\Scale[0.9]{
  A_Y :=A_Z-B_ZD_Z^{-1}C_Z,
  B_Y :=B_ZD_Z^{-1},
  C_Y :=-D_Z^{-1}C_Z,   D_Y :=D_Z^{-1}.
}
\] 
\end{definition}
\noindent
It can be easily noted that if the system $\Sigma_Z:(A_Z,B_Z,C_Z,D_Z)$, has
symmetric state-space  realization with $A_Z=A_Z^T, B_Z= C^T_Z, D_Z=D^T_Z$ with $D_Z$-invertible, then the inverse system $\Sigma_Y$ has also symmetric state-space  realization but with $B_Y=- C_Y^T$. The poles of $\Sigma_Z$ are the zeros of the system $\Sigma_Y$ and vice-versa. 
It is interesting to note that the inverse systems share the same set of spectral zeros i.e. the spectral zeros are invariant to i/o partition.

\begin{lemma}\label{lem:spec-zero invariant inverse}
Consider a MIMO system $\Sigma_Z:$ $ (A_Z,B_Z,C_Z,D_Z)$ with its
inverse system $\Sigma_Y:$ $ (A_Y,B_Y,C_Y,D_Y)$ as
in Definition~\ref{defn:inverseG:state-space}.
Then, the Hamiltonian matrix with respect to the passivity supply rate $u^T y$
for the system $\Sigma_Z$ and its inverse $\Sigma_Y$ are the same.
Consequently, the spectral-zeros of $\Sigma_Z$ and
$\Sigma_Y$ are the same.
\end{lemma}
In view of the spectral zeros being eigenvalues of the Hamiltonian matrix $H$,
we denote by $\SZP(s)$ the polynomial whose roots, counted with multiplicity,
are the spectral zeros, both stable and anti-stable. $\SZP(s)$ is nothing
but the characteristic polynomial of $H$.

\begin{proof}
  The Hamiltonian matrix $H_Z$ of the system $\Sigma_Z:(A_Z,B_Z,C_Z,D_Z)$ is:
  \begin{equation*} \Scale[0.9]{
    \begin{array}{cl}
     H_Z \!&=\! \begin{bmatrix}
     A_Z-B_Z(D_Z+D_Z^T)^{-1}C_Z \! \!&B_Z(D_Z+D_Z^T)^{-1}B_Z^T         \\
     -C_Z^T(D_Z+D_Z^T)^{-1}C_Z \!\! & -(A_Z-B_Z(D_Z+D_Z^T)^{-1}C_Z)^T        
    \end{bmatrix}\\
    &=:\begin{bmatrix}
    P_Z &Q_Z\\ R_Z &-P_Z^T 
    \end{bmatrix}, \mbox{ say, with blocks defined appropriately}.
    \end{array}}
  \end{equation*} 
  The Hamiltonian matrix $H_Y$ of the inverse system is 
 \begin{equation*} \Scale[0.85]{
   \begin{array}{cl} 
  H_Y \! &\!=\!\begin{bmatrix}
  A_Y-B_Y(D_Y+D_Y^T)^{-1}C_Y \! \!\! &B_Y(D_Y+D_Y^T)^{-1}B_Y^T         \\
  -C_Y^T(D_Y+D_Y^T)^{-1}C_Y \! \!\! &-(A_Y-B_Y(D_Y+D_Y^T)^{-1}C_Y)^T        
  \end{bmatrix} \\
  &\!=:\begin{bmatrix}
  P_Y &Q_Y\\ R_Y &-P_Y^T 
   \end{bmatrix}, \mbox{ say, with blocks defined appropriately}.
\end{array} }
  \end{equation*} 
Notice that $Q_Y=B_ZD_Z^{-1} (D_Z^{-1}+D_Z^{-T})^{-1}D_Z^{-T}B_Z^T=Q_Z$ and $R_Y=-C_Z^T D_Z^{-T}(D_Z^{-1}+D_Z^{-T})^{-1} D_Z^{-1}C_Z=R_Z$. Further,
\begin{align*}
P_Y&=A_Z-B_ZD_Z^{-1}C_Z + B_ZD_Z^{-1}(D_Z^{-1}+D_Z^{-T})^{-1}D_Z^{-1}C_Z \\
   &=A_Z-B_Z[D_Z^{-1} - D_Z^{-1}(D_Z^{-1}+D_Z^{-T})^{-1}D_Z^{-1}]C_Z ~.
\end{align*}
We now use the Matrix Inverse Lemma (also called the Sherman Morrison Woodbury
formula~\cite[Theorem 0.7.4]{HornJohn2012}), which states that for
nonsingular square matrices $A$ and $R$ (of possibly different sizes), the following holds:
\begin{equation*}\label{eqn:woodbury formula}
   (A + XRY)^{-1}= A^{-1}-A^{-1}X(R^{-1} + Y A^{-1}X)^{-1}Y A^{-1}
\end{equation*}
with $X, Y$ and $R$ of appropriate dimensions.
Using the above relation expand $(D_Z+D_Z^T)^{-1}$, by replacing
$A=D_Z, X=Y=I_n$ and $R=D_Z^T$, to get
\begin{equation*}
  (D_Z+D_Z^T)^{-1}=D_Z^{-1} - D_Z^{-1}(D_Z^{-1}+D_Z^{-T})^{-1}D_Z^{-1} ~.
\end{equation*}
  Applying the above equality, write $P_Y$ as:
\begin{equation*}
  P_Y= A_Z-B_Z(D_Z+D_Z^T)^{-1}C_Zv =P_Z ~.
\end{equation*}
Therefore we get $H_Z=H_Y$.
  Hence the system $\Sigma_Z$ and its inverse system  $\Sigma_Y$ have the same Hamiltonian matrix. As a result spectral-zeros of both the systems $\Sigma_Z$ and $\Sigma_Y$ are identical.
\end{proof}
Obvious from the above lemma and its proof is that
the ARE and its solutions are also identical for a system and its inverse-system, i.e.
these properties are invariant of the i/o partition. 
As a fallout, it can be easily seen that, with $Z_1$ and $Z_2$ as two
arbitrary SISO transfer functions, the spectral zeros of
the following MIMO systems $(G_i)$ are all the same set:
\begin{equation*}
\Scale[0.88]{G_1=\begin{bmatrix}
Z_1 \!\! &0\\0 \!\!&Z_2
\end{bmatrix},G_2=\begin{bmatrix}
Z_1^{-1} \!\!&0\\0 \!\!&Z_2
\end{bmatrix},G_3=\begin{bmatrix}
Z_1 \!\!&0\\0 \!\!&Z_2^{-1}
\end{bmatrix}, G_4=\begin{bmatrix}
Z_1^{-1} \!\!&0\\0 \!\!&Z_2^{-1}
\end{bmatrix}}.
\end{equation*} 
This observation can be used to illustrate that the ZIP property
presented for the SISO case in Theorem~\ref{thm:real:spectral:zeros_siso}
would not get extended to MIMO systems in an obvious way. Below
is a more specific and simple counterexample:
a decoupled MIMO system given by transfer matrix $G(s)$:
\begin{equation*}
G(s)=\begin{bmatrix}
\frac{(s+1)(s+5)}{(s+3)(s+7)} &&0 \\
0 &&\frac{(s+2)(s+6)}{(s+4)(s+8)}
\end{bmatrix}~.
\end{equation*} 
$G(s)$ is made up of two SISO transfer functions with ZIP property. The poles, zeros and spectral zeros of $G(s)$ are:
\begin{equation*}
\begin{array}{ll}
\text{system-poles} \!\!& \!\!:p_1=-8,\; p_2=-7,\;p_3=-4,\;p_4=-3,\\
\text{system-zeros}  \!\!& \!\!: z_1=-6,\;z_2=-5,\;z_3=-2,\;z_4=-1,\\
\text{spectral-zeros}  \!\!& \!\!:\MU_1=\pm6.5,\;\MU_2=\pm5.5,\;\MU_3=\pm2.9,\;\MU_4=\pm1.9 ~.
\end{array}
\end{equation*}
Therefore, the MIMO system $G(s)$ does not exhibit the ZIP property as there are no system-poles between system-zero pairs $z_1/z_2$ and $z_3/z_4$ while
two system-poles $p_3$ and $p_4$ lie between system-zero pair $z_2/z_3$.
However, it is interesting to observe that the each stable
spectral zero $\MU_i$ occurs between a system-pole/zero pair.

Having seen a MIMO example of the absence of the ZIP property, we now move towards
a subset of MIMO systems which we study further and prove results regarding
interlacing of poles/system-zeros and spectral zeros. We first prove
that spectral zeros are real for the class of MIMO systems admitting a
symmetric state-space realization.
\begin{theo} \label{thm:real:spectral:zeros_mimo}
A strictly passive MIMO system that admits a
symmetric state-space realization has all spectral zeros real.
\end{theo}

\begin{proof}
Consider a strictly passive MIMO system $\Sigma$ with symmetric state-space
realization $ (A=A^T, B=C^T, D=D^T)$. (The proof for
$B=-C^T$ is identical and is not reproduced here.)
The Hamiltonian matrix $H$ is as follows:
  \begin{align*}
  H= \begin{bmatrix}
  A-B(D+D^T)^{-1}C & B(D+D^T)^{-1  }B^T         \\
  -C^T(D+D^T)^{-1}C & -(A-B(D+D^T)^{-1}C)^T        
  \end{bmatrix} ~.
  \end{align*}
Let $P:=B(D+D^T)^{-1  }B^T=C^T(D+D^T)^{-1}C$. Since $A=A^T$ and $P=P^T$, the Hamiltonian matrix $H$ can be represented as
  \begin{equation*}
  H= \begin{bmatrix}
  A-P & P         \\
  -P & -A+P        
  \end{bmatrix}.
  \end{equation*}
  Using a similarity transformation of the
Hamiltonian matrix $(T^{-1} HT)$ where  $T=\big[\begin{smallmatrix}I &&0\\I &&I\end{smallmatrix}\big]$ and $T^{-1}=\big[\begin{smallmatrix}I &&0\\-I &&I\end{smallmatrix}\big]$
we get $H=$
\[
\begin{bmatrix}I & 0\\-I &I\end{bmatrix} \begin{bmatrix}
  A-P & P         \\
  -P & -A+P        
  \end{bmatrix} \begin{bmatrix}I &0\\I &I\end{bmatrix}
=\begin{bmatrix}
  A & P         \\
  -2A & -A        
  \end{bmatrix}.
\]
Computing the square of the Hamiltonian matrix, we get
\begin{equation} \label{eqn: Square Hamiltonian}
  H^2\! 
  =\!\begin{bmatrix}
  A\!\! &\!\! P         \\
  -2A\!\! &\!\! -A        
  \end{bmatrix} \begin{bmatrix}
  A\!\! &\!\! P         \\
  -2A\!\! &\!\! -A        
  \end{bmatrix}\! =\!
\begin{bmatrix}
  A^2-2PA\!\! &\!\!AP-PA          \\
  0\!\! &\!\! A^2-2AP        
  \end{bmatrix}.
\end{equation}
  
  Now since the block-diagonal entries satisfy: $(A^2-2AP)^T=(A^2-2PA)$, 
eigenvalues of $H^2$ are same as the eigenvalues of $(A^2-2AP)$, but with
multiplicities doubled.
  
  Applying, similarity transform of $(A^2-2AP)$ using $T:=\sqrt{-A~}$, the
square-root\footnote{For a symmetric and positive definite matrix $P$, we
    define $\sqrt{P}$ as the unique symmetric and positive definite matrix
    that satisfies $(\sqrt{P})^2 =P$ and denote its
    inverse as$\sqrt{P}^{~-1}=P^{~-\frac{1}{2}}$.  } we get:
  \begin{align*}
  A^2-2AP &= \sqrt{-A~}^{~-1} (A^2-2AP)\sqrt{-A~} \\
  &=A^2-2\sqrt{-A~} P \sqrt{-A~} ~.
  \end{align*}
  Now, $\sqrt{-A~}P\sqrt{-A~}$ is symmetric and hence so is $(A^2-2\sqrt{-A~} P \sqrt{-A~})$.
  Therefore, $(A^2-2AP)$ has real eigenvalues i.e. $H^2$ has real eigenvalues.
  Also, we know that the eigenvalues of $H^2$ are squared eigenvalues of $H$ and 
since the system is strictly passive, $H$ has real eigenvalues.
\end{proof}

It is important to note that,
in order for a system to exhibit the ZIP property,
it is essential for the system to have distinct poles.
In the SISO case, a symmetric state-space  realizable system which is controllable and observable automatically dictates that the system-poles are distinct and hence no additional assumptions are required. But in the MIMO case a symmetric state-space  realizable system which is controllable and observable does not guarantee distinct system-poles. 
The following lemma helps in the proof of the main MIMO results.
\begin{lemma} \label{lem:A plus P}
  Consider symmetric matrices $P,M \in \Rnn$ with $P$ having distinct eigenvalues and
  $M$ positive semidefinite symmetric matrix of rank $r$ with $(r<n)$ and
  with the eigenvalues of each matrix ordered as in
  Eqn.~\eqref{eqn: eigenvalues ordering}.
  Suppose the largest eigenvalue of $M$ is at most the minimum difference between any two eigenvalues of $P$ i.e.
  \begin{equation} \label{eqn:largest eigenvalueM}
  \lambda_n(M) \leqslant
  \hspace*{3mm} \underset{\displaystyle i=1,\dots,n-1}{\min}
  (\lambda_{i+1}(P)-\lambda_i(P) ). 
  \end{equation}
  Then, the following statements hold.  
\begin{enumerate}
\item The eigenvalues of $P$ and $(P+M)$ interlace, i.e.\footnote[5]{Note
         that amongst the two inequalities
         within Eqn.~\eqref{eqn:A plus P lemma},
          index $i$ varies from $1$ to $n$ in the first, while varies
         from $1$ to $n-1$ in the second. This slight abuse of indexing notation
         helps convey the interlacing property and avoids repetition. 
         Same has been pursued at other similar inequalities also.}
\begin{equation} \label{eqn:A plus P lemma}
     \hspace*{-5mm} \Scale[0.95]{\lambda_{i}(P)\leqslant \lambda_{i}(P+M) \leqslant \lambda_{i+1}(P) \;\text{ for 
       each 
         } i=1,2,\cdots,n.  }
    \end{equation}
\item Further, if $Mx\neq0$ for every eigenvector $x$ of $P$ and the inequality is strict in Eqn.~\eqref{eqn:largest eigenvalueM}, then
    \begin{enumerate} \itemindent -2mm
      \item the eigenvalues of $(P+M)$ are distinct
      \item the eigenvalues of $P$ and $P\!+\!M$ interlace {\bf strictly}:
      \begin{equation} \label{eqn:strictAplusPlemma}
\hspace*{-5mm}  \Scale[0.9]{ \lambda_{i}(P)< \lambda_{i}(P+M) < \lambda_{i+1}(P) \;\text{ for 
     each } i=1,2,\cdots,n. }
      \end{equation}
    \end{enumerate}  
  \end{enumerate}   
\end{lemma}

\begin{proof}
  Utilizing the Weyl's inequality 
  theorem (see \cite[Theorem 4.3.1]{HornJohn2012})\footnote[6]{
    \textbf{\bf Weyl's inequality theorem:} Let $A, B\in \Rnn$ 
    be symmetric and suppose the respective eigenvalues
    of $A, B$ and $A + B$ be $\{\lambda_i(A)\}_{i=1}^n$, $\{\lambda_i(B)\}_{i=1}^n$ and $\{\lambda_i(A+B)\}_{i=1}^n$ each
    algebraically ordered in non-increasing order such that:
    \begin{equation*} 
    \lambda_{\min}=\lambda_1 \leqslant \lambda_2 \leqslant \cdots \leqslant \lambda_{n-1} \leqslant \lambda_n=\lambda_{\max}.  
    \end{equation*} 
    Then
    \begin{align*}
    \lambda_i(A+B) \leqslant \lambda_{i+j}(A) + \lambda_{n-j}(B)\;\; j=0,1,\ldots,(n-i)
    \end{align*}
    for each $i=1,\ldots,n$, with equality for some pair $(i,j)$ if and only if there is a nonzero vector $x$
    such that $Ax = \lambda_{i+j}(A)x,\; Bx = \lambda_{n-j} (B)x, \text{ and } (A + B)x = \lambda_i (A + B)x$.
    Also,
    \begin{equation*}
    \lambda_{i-j+1}(A)+\lambda_j(B)\leqslant\lambda_i(A+B)\;\;\; j=1,2,\ldots,i
    \end{equation*}
    for each $i=1,\ldots,n$, with equality for some pair $(i,j)$ if and only if there is
    a nonzero
    vector $x$ such that $Ax = \lambda_{i−j+1}(A)x,\; Bx = \lambda_j (B)x,
    \text{ and } (A + B)x = \lambda_i (A + B)x$. If
    $A$ and $B$ have no common eigenvector, then every inequality
    in the above equation is a strict
    inequality.  } 
  we  write:
  \begin{align}
  \!\!\! &\!\!\!\lambda_i(P+M) \leqslant \lambda_{i+j}(P)\!\! + \lambda_{n-j}(M)\;\! \mbox{ for } j\!=\!0,1,\!\ldots\!,\!(n-i)\! \label{eqn:Weyl1}\\
  \!\!\! &\!\!\!\lambda_{i-j+1}(P)+\lambda_j(M)\leqslant\lambda_i(P+M)\;\;\;  \mbox{ for } j=1,2,\ldots,i ~.\label{eqn:Weyl2}
  \end{align}
  Using the Eqn.~\eqref{eqn:Weyl1} for $j=0$ we
obtain:
  \begin{equation} \label{eqn:Weylpositive1}
  \lambda_i(P+M)\leqslant \lambda_i(P)+\lambda_n(M) \; \text{ for } i=1,2,\cdots,n ~.
  \end{equation} 
  
  Similarly, from the Eqn.~\eqref{eqn:Weyl2} for $j=1$, one can write:
  \begin{equation*}
  \lambda_{i}(P)+\lambda_1(M)\leqslant \lambda_i(P+M) ~.
  \end{equation*}
  Since rank of $M$ is $r<n$, $\lambda_1(M)=0$. Therefore, we write:
  \begin{equation} \label{eqn:Weyl positive2}
  \lambda_i(P) \leqslant \lambda_{i}(P+M) \; \text{ for all } i=1,2,\cdots,n ~.
  \end{equation}
  
  Now combining the above Eqns.~\eqref{eqn:Weylpositive1} and~\eqref{eqn:Weyl positive2} we write:
  \begin{equation*}
  \lambda_{i}(P) \leqslant \lambda_i(P+M)\leqslant \lambda_i(P)+\lambda_n(M) ~.
  \end{equation*}
  Therefore, if
  \begin{equation*}
  \lambda_i(P) + \lambda_n(M) \leqslant \lambda_{i+1}(P) \; \text{ for } i=1,2,\ldots,(n-1),
  \end{equation*}
  then one can write:
  \begin{equation} \label{eqn:Weyl positive 3}
  \lambda_{i}(P) \leqslant \lambda_i(P+M)\leqslant \lambda_{i+1}(P) \;\text{ for } i\!=\!1,2,\!\ldots,\!(n-1)\!
  \end{equation}
  i.e.
  \begin{equation*}
  \lambda_1(\!P\!)\leqslant \lambda_1(\!P\!+\!M\!)\leqslant \lambda_2(\!P\!) \leqslant \lambda_2(\!P\!+\!M\!)\leqslant\!\! \ldots\!\! \leqslant\lambda_n(\!P\!) \leqslant \lambda_n(\!P\!+\!M\!)
  \end{equation*}
  This proves Statement~$1$ of Lemma~\ref{lem:A plus P}.\\
  It can be easily seen that if for any eigenvector $x$ of $P$ s.t. $Px=\lambda_i x$, $Mx=0$ then $(P+M)x=Px=\lambda_i x$; the corresponding inequality in Eqn.~\eqref{eqn:Weyl positive2} becomes an equality i.e $\lambda_i(P) = \lambda_{i}(P+M)$.\\
  Therefore, if for every eigenvector $x$ of $P$, $Mx\neq0$ then every inequality in Eqn.~\eqref{eqn:Weyl positive2} becomes  strict inequality and therefore we get:
  \begin{equation} \label{eqn:Weyl strict positive}
  \lambda_i(P) < \lambda_{i}(P+M) \; \text{ for all } i=1,2,\cdots,n ~.
  \end{equation}
  Further, assuming
  \begin{equation*} 
  \lambda_n(M) <
  \hspace*{3mm} \underset{\displaystyle i=1,\dots,n-1}{\min}
  (\lambda_{i+1}(P)-\lambda_i(P) ), 
  \end{equation*} and combining this with Eqn.~\eqref{eqn:Weylpositive1} we get
  \begin{equation*} 
  \begin{array}{cl}
  &\lambda_i(P+M)\leqslant \lambda_i(P)+\lambda_n(M)\; <\; \lambda_{i+1}(P), \\
  \implies &\lambda_i(P+M)< \lambda_{i+1}(P) \;\text{ for } i=1,2,\cdots,(n-1)~.
  \end{array}
  \end{equation*} 
Therefore combining with Eqn.~\eqref{eqn:Weyl strict positive} we get:
$ \lambda_i(P)<\lambda_i(P+M)< \lambda_{i+1}(P) \mbox{ for } i=1,2,\ldots,n$, i.e.
\begin{equation*} \label{eqn:Weyl positive 4}
   \Scale[0.9]{
    \!\!\!\lambda_{1}(P)<\lambda_{1}(\!P\!+\!M\!) <\lambda_{2}(P)<\!\ldots\!<\lambda_{n-1}(\!P\!+\!M\!) <\lambda_{n}(P)< \lambda_{n}(\!P\!+\!M\!) \!\!
  }
\end{equation*} 
  Since all the eigenvalues of $P$ are distinct, from the above equation 
  it is easily seen that the eigenvalues of $P+M$ are also distinct. 
\end{proof}

Following the above Lemma~\ref{lem:A plus P}, the relation between eigenvalues of $P$ and $(P-M)$ is given as:
  \begin{equation} \label{eqn:A minus P lemma}
 \lambda_{i}(P-M) < \lambda_i(P) < \lambda_{i+1}(P-M) \;\;\text{ for } i=1,2,\!\ldots\!,n \!
 \end{equation}
if, $\lambda_n(M) < \underset{\displaystyle i=1,\dots,n-1}{\min}(\lambda_{i+1}(P)-\lambda_i(P) )$ and for every eigenvector $x$ of $P$, $Mx\neq0$.

\begin{lemma} \label{lem:product P(P+M)}
  Suppose $P$ and $M\in\Rnn$ are both symmetric, let $P$ be positive definite and
  $M$ be singular and positive semidefinite. Then, the following hold.
  \begin{enumerate}
    \item The set of eigenvalues of the products of $P(P+M)$ and $(P+M)P$ coincide, i.e.
    $ 
    \lambda(P(P+M)) 
    = \lambda((P+M)P). 
    $ 
    \item Eigenvalues of the product $P(P+M)$ are real.
    \item Eigenvalues of the product $P(P+M)$ lie between
    the eigenvalues of $P^2$ and $(P+M)^2$. 
    \begin{equation} \label{eq:interlacing:product}\Scale[0.9]{
    \!\!\!\lambda_i^2(P) \leqslant \lambda_i(P(P+M)) \leqslant \lambda_i^2(P+M) \;\; \text{ for } i\!=\!1,2,\!\ldots\!,n.\!\!
     }\end{equation} 
    \item Suppose for every eigenvector $x$ of $P$, we have $Mx \neq 0$.
    Then each of the inequalities in Eqn.~\eqref{eq:interlacing:product}
    are strict, i.e.
    \begin{equation*}
    \lambda_i^2(P) < \lambda_i(P(P+M)) < \lambda_i^2((P+M))  \text{ for } i=1,2,\ldots,n.
    \end{equation*}
  \end{enumerate} 
\end{lemma}
\begin{proof}
Consider $P$ and $M\in\Rnn$ with $P$- symmetric positive definite and $M$-symmetric positive semi-definite.
Now 
\[
  (P(P+M))^T=(P+M)^TP^T=(P+M)P ~.
\]
Therefore, eigenvalues of $P(P+M)$ and $(P+M)P$ coincide, i.e. $\lambda(P(P+M))= \lambda((P+M)P)$. This proves statement$(1)$ of the lemma.
Now,
\[
  \lambda(P(P+M))= \lambda(P^2+PM) ~.
\]
  $P$ is symmetric positive definite, we define $T=\sqrt{P~}$ and do similarity transform of $P^2+PM$:
  \begin{equation*}
  \begin{array}{cl}
  \lambda(P(P+M)) &=\lambda(T^{-1}(P^2+PM)T) \\
  &=\lambda(\sqrt{P~}^{~-1}P^2\sqrt{P~}+\sqrt{P~}^{~-1}PM\sqrt{P~})\\
  &=\lambda(P^2+\sqrt{P~}M\sqrt{P~}) ~.
  \end{array}
  \end{equation*}
  Since $P^2$ is symmetric positive definite and $\sqrt{P~}M\sqrt{P~}$- is symmetric positive semi-definite with rank $r<n$, therefore $(P^2+\sqrt{P~}M\sqrt{P~})$ is symmetric positive definite and hence $P(P+M)$ has real eigenvalues. This proves statement $(2)$.
  
  Now, applying Lemma~\ref{lem:A plus P} and using Eqn.~\eqref{eqn:A plus P lemma} we get:
  \begin{align} \label{eqn:A2 part 1}
  &\lambda_i(P^2) \leqslant \lambda_i(P(P+M))  \nonumber \\
  \implies &\lambda_i^2(P) \leqslant \lambda_i(P(P+M)) ~.
  \end{align}
  Now if we denote $Q:=P+M$, $Q$ is symmetric positive definite and we get:
  \begin{equation*}
  \lambda(P(P+M))=\lambda((Q-M)Q)=\lambda(Q^2-MQ) ~.
  \end{equation*}
  $Q$ is symmetric positive definite, we define $T^{-1}=\sqrt{Q~}$ and do similarity transform of $Q^2-MQ$:
  \begin{equation*}
  \begin{array}{cl}
  \lambda(P(P+M)) &=\lambda(Q^2-MQ)=\lambda(T^{-1}(Q^2-MQ)T) \\
  &=\lambda(\sqrt{Q~}Q^2\sqrt{Q~}^{~-1}-\sqrt{Q~}MQ\sqrt{Q~}^{~-1})\\
  &=\lambda(Q^2-\sqrt{Q~}M\sqrt{Q~}) ~.
  \end{array}
  \end{equation*}
  Again $Q^2$ is symmetric positive definite and $\sqrt{Q~}M\sqrt{Q~}$ is
   symmetric positive semi-definite with rank $r<n$, applying Lemma~\ref{lem:A plus P} and using Eqn.~\eqref{eqn:A minus P lemma} we get:
  \begin{align} \label{eqn:A2 part 2}
  & \lambda_i(P(P+M)) \leqslant \lambda_i(Q^2) \nonumber \\
  \implies &\lambda_i(P(P+M)) \leqslant \lambda_i^2(P+M)  ~.
  \end{align}
  Therefore, combining Eqns.~\eqref{eqn:A2 part 1} and~\eqref{eqn:A2 part 2} we get:
  \[
  \lambda_i^2(P) \leqslant \lambda_i(P(P+M)) \leqslant \lambda_i^2(P+M) \;\; \text{ for each }\; i=1,2,\ldots,n.
  \]
  If for every eigenvector $x$ of $P$, we have $Mx \neq 0$, this implies that $\sqrt{P~}M\sqrt{P~}x \neq 0$ as $\sqrt{P}$ has the same set of eigenvectors as $P$. Similarly for every eigenvector $x$ of $Q$, $\sqrt{Q~}M\sqrt{Q~}x \neq 0$.
Therefore, from Lemma~\ref{lem:A plus P}, we get the required inequality
\[
  \lambda_i^2(P) < \lambda_i(P(P+M)) < \lambda_i^2((P+M)) 
\]
for each $i=1,2,\ldots,n$, thus completing the proof.
\end{proof}
For a strictly passive system $\Sigma_Z:(A_Z,B_Z,C_Z,D_Z)$ in symmetric state-space realization $A_Z=A_Z^T, B_Z=C^T_Z, D_Z=D^T_Z$, the poles of the system are eigenvalues of  $A_Z$ and the zeros of the system are defined as the eigenvalues of $A_Y=A_Z-B_ZD_Z^{-1}C_Z=A_Z-B_ZD_Z^{-1}B^T_Z$. We are interested in characterizing the conditions for which the system-poles and zeros interlace. 


This leads to our next result, but before that we need to define the difference between two consecutive system-poles of the ordered set $\{p_i(\Sigma)\}_{i=1}^n$ by $\nu(\Sigma)$:
\begin{equation*}
\nu_i(\Sigma):=p_{i+1}(\Sigma)-p_i(\Sigma)  \;\; \text{ for }\;\;i=1,2,\ldots,(n-1)~.
\end{equation*} 
and denote the minimum difference between the system-poles $p(\Sigma)$ as 
\begin{equation*}
\nu_{\min}=\min_{i=1:n-1}\nu(\Sigma) ~.
\end{equation*}

\begin{theo} \label{thm:ZIP MIMO}
Consider a strictly passive controllable MIMO system $\Sigma$ that admits a symmetric state-space 
realization with distinct system-poles.
If the minimum difference between the system-poles is greater than
the largest eigenvalue of $(B D^{-1}B^T) $, i.e.
  \begin{equation}
  \nu_{\min}(\Sigma) > \lambda_{\max}(B D^{-1}B^T).
  \end{equation}
Then,
  the system-poles and system-zeros interlace strictly:
  \begin{equation*}
    \begin{array}{lcl}
      \text{for }\;B=C^T  \!\!&\!\!: z_1<p_1<z_2<p_2<z_3<\cdots< p_{n-1}<z_n<p_n,\\
      \text{for }\;B=-C^T \!\!&\!\!: p_1<z_1<p_2<z_2<p_3<\cdots< z_{n-1}<p_n<z_n.
    \end{array} 
  \end{equation*}
\end{theo}

\begin{proof}
We prove Theorem~\ref{thm:ZIP MIMO} for just the case of $B=C^T$ since proof for the other case $B=-C^T$ is
analogous.
Consider a strictly passive controllable MIMO system $\Sigma$ with symmetric state-space  realization $A=A^T, B=C^T, D=D^T$ and we use that the system-poles are distinct.
Next, the poles of the system $\Sigma$ are $p(\Sigma)=\lambda(A)$ and system-zeros are $z(\Sigma)=\lambda(A-BD^{-1}B^T)$.
Define $P:=BD^{-1}B^T$ and express the system-zeros as: $z(\Sigma)=\lambda(A-P)$.
The minimum difference between the system-poles is greater than the largest eigenvalue of $B D^{-1}B^T$: this means
\begin{equation*}
\nu_{\min} > \lambda_{\max}(B D^{-1} B^T)= \lambda_{\max}(P).
\end{equation*}
 As the system is $(A,B)$ controllable, therefore from the Popov-Belevitch-Hautus
 (PBH) test for controllability, we get that for every left-eigenvector $w_i$ of $A$ s.t. $w_i^TA=\lambda_i w_i^T$,
 \[
 w_i^TB \neq 0 ~.
 \]
 Since, $A=A^T$, the left and right eigenvector are the same, we get that for every eigenvector $x_i$ of $A$, $B^Tx_i \neq 0$. As the system is strictly passive $D$ is positive definite symmetric matrix, therefore we write:
 \begin{equation}\label{eqn:Pxi neq 0}
   Px_i=B D^{-1} B^T x_i \neq 0 \;\; \text{ for all } i=1,2,\ldots,n ~.
 \end{equation} 
Therefore, utilizing Lemma~\ref{lem:A plus P} and Eqn~\eqref{eqn:A minus P lemma}, we get
 $z_1<p_1<z_2<\ldots< p_{n-1}<z_n<p_n$, thus completing the proof.
\end{proof}

\begin{lemma} \label{lem:sufficiently-large-D-for-Z}
Consider a strictly passive controllable MIMO
system $\Sigma_Z$ with distinct system-poles which admits a symmetric state-space  realization $A_Z=A_Z^T, B_Z=C^T_Z, D_Z=D^T_Z$.  
Suppose the feed-through term is a scaled version of a fixed matrix $D >0$, 
i.e. $D_Z=\eta D,\;\eta \in \R_+$. Then, for sufficiently large $\eta$, 
the system-poles/zeros are interlaced strictly.
\end{lemma}
\begin{proof}
  Consider a strictly passive controllable MIMO system $\Sigma_Z:(A_Z \in\Rnn,B_Z \in \Rnp,C_Z \in \Rmn,D_Z \in \Rmp)$ with ($m=p$ and $m<n)$ in symmetric state-space  realization $A_Z=A_Z^T, B_Z=C^T_Z, D_Z=D^T_Z>0$
  We consider a scaling factor $\eta$ for the feed-through matrix $D_Z$ such that $D_Z=\eta D$ with $\eta >0$.
  Now the system-poles and system-zeros are given as:
  \begin{equation*}
    \begin{array}{cl}
    p(\Sigma_Z)&:=\lambda(A_Z),\\
    z(\Sigma_Z)&:=\lambda(A_Z-\frac{1}{\eta}B_zD^{-1}B_Z^T)~.
    \end{array}
  \end{equation*}
  From Theorem~\ref{thm:ZIP MIMO} we get that for the system $\Sigma_Z$, the system-zeros and poles
are interlaced if $\nu_{\min}(\Sigma) > \lambda_{\max}(B_Z D_Z^{-1}B_Z^T)$, which
is 
satisfied if $ \nu_{\min}(\Sigma) > \frac{1}{\eta}\lambda_{\max}(B_Z D^{-1}B_Z^T) $.
Thus, for $\eta$ sufficiently large, the above inequalities 
are satisfied, and hence the system-poles and zeros are interlaced strictly.
\end{proof}

\noindent
Though the feed-through matrices of 
transfer matrices $\Sigma_Y$ and $\Sigma_Z$ are inverses of each other, thus suggesting
that the interlacing conclusion on the system poles and zeros would be obtained for $\Sigma_Y$
for a sufficiently {\em small} (and positive definite) $D_Y$, counter-intuitively,
interlacing happens again for a sufficiently \emph{large} $D_Y$: hence we record
it as a lemma.
\begin{lemma}   \label{lem:sufficiently-large-D-for-Y}
Consider a strictly passive controllable MIMO
system $\Sigma_Y$ admitting a symmetric state-space  realization
 $A_Y=A_Y^T, B_Y=-C^T_Y, D_Y=D^T_Y>0$ and with distinct poles.  
Suppose the feed-through matrix $D_Y$ is scaled
as $D_Y=\eta D,\;\eta \in \R_+$.
Then, for a sufficiently large $\eta$, the system-poles/zeros are
interlaced strictly.
\end{lemma}
The proof of Lemma~\ref{lem:sufficiently-large-D-for-Y} is omitted
as it closely follows the proof of Lemma~\ref{lem:sufficiently-large-D-for-Z}.
The following two theorems pertain to system-pole/zero and spectral zero
interlacing for MIMO systems are amongst
the main results of this section.

\begin{theo}\label{thm:SZ with ZIP MIMO}
Consider a strictly passive controllable MIMO system $\Sigma$ that admits a symmetric 
state-space realization and exhibits ZIP property, i.e. after ordering and indexing
the poles/zeros as
in Eqn.~\eqref{eqn: eigenvalues ordering}, we have:
\begin{equation*}
\begin{array}{lcl}
  \text{for }\;B=C^T  \!\!&\!\!: z_1<p_1<z_2<p_2<z_3<\cdots< p_{n-1}<z_n<p_n,\\
  \text{for }\;B=-C^T \!\!&\!\!: p_1<z_1<p_2<z_2<p_3<\cdots< z_{n-1}<p_n<z_n.
\end{array} 
\end{equation*}
 Then, the stable spectral zeros of the system $\mu(\Sigma)^-$
are also interlaced strictly
between the pair of system poles $p(\Sigma)$ and zeros $z(\Sigma)$:  
\begin{equation*}\Scale[0.90]{
    \begin{array}{ll}
 \!\!\!    \!\!\!\text{ for $B=C^T$}  \!\!&\!\!:z_1<\MU_1<p_1<z_2<\MU_2<p_2<
   \Scale[0.9]{\cdots} <p_{n-1}<z_n<\MU_n<p_n,\\
 \!\!\!    \!\!\!\text{ for $B=$-$C^T$}  \!\!&\!\!:p_1<\MU_1<z_1\;<p_2<\MU_2<z_2<
   \Scale[0.9]{\cdots} < z_{n-1}<p_n<\MU_n<z_n.
   \end{array}
   }
\end{equation*}
\end{theo}

\noindent
The proof of Theorem~\ref{thm:SZ with ZIP MIMO} is presented within
the proof of Theorem~\ref{thm:SZ betwen Pole Zero pair} as the former is a
special case of the latter.
The next result states 
that, even if the poles and system zeros of MIMO systems (that admit a symmetric 
state-space realization, etc) are not interlaced, the spectral zeros
still occur between the appropriate system-pole/zero pair. 

\begin{theo} \label{thm:SZ betwen Pole Zero pair}
For a strictly passive controllable MIMO system $\Sigma$ that admits symmetric state-space 
realization, each stable spectral zero
occurs between a system pole-zero pair. More precisely, suppose
the system-poles $p_i(\Sigma)$, 
system-zeros $z_i(\Sigma)$ and the stable spectral zeros $\mu_i(\Sigma)^-$ are ordered and indexed 
as in Eqn.~\eqref{eqn: eigenvalues ordering}. Then:
  \begin{equation}
    \begin{array}{cccl}
    z_i(\Sigma)&\leqslant \MU_i(\Sigma)^-&\leqslant p_i(\Sigma)  &\text{ for } B= + C^T~, \\
    p_i(\Sigma)&\leqslant \MU_i(\Sigma)^-&\leqslant z_i(\Sigma)  &\text{ for } B=-C^T ~.
    \end{array}
  \end{equation}
\end{theo}

\begin{proof}
  Consider a strictly passive MIMO system $\Sigma$ which admits a controllable
symmetric state-space  realization $(A=A^T, B=\pm C^T, D=D^T)$. 
We prove below only for the case $B=C^T$, since the proof
of the case $B=-C^T$ follows closely.

We know that the spectral zeros  $\MU(\Sigma)$ are the eigenvalues of the Hamiltonian matrix $H(\Sigma)$ with respect to the passivity supply rate. 
  From the Eqn.~\eqref{eqn: Square Hamiltonian} the eigenvalues of square of Hamiltonian matrix are expressed as:
 {\small \begin{align*} 
  \lambda(H^2(\Sigma)) &= \lambda(A^2 - 2 B(D+D^T)^{-1}B^TA) = \lambda(A^2-BD^{-1}B^T A)\\&= \lambda((-A)^2+BD^{-1}B^T(-A))~.
  \end{align*} }
Define $P:=-A$ and $M:=BD^{-1}B^T$ and the set of
stable spectral zeros as $\MU(\Sigma)^+$. The above equation is nothing but
\[
\MU^2(\Sigma)^+=\lambda((P+M)P).
\]
We order and index the set of stable spectral zeros $\MU^(\Sigma)^+$ and eigenvalue sets  $\lambda(P)$ and $\lambda(P+M)$ as per Eqn.~\eqref{eqn: eigenvalues ordering}.
As $P$ is symmetric and positive definite and $M$ is symmetric and
positive semi-definite,
utilizing Lemma~\ref{lem:product P(P+M)} we get that for
$i=1,2,\ldots,n$:
\begin{equation}\label{eqn:P(P+M) eigenvalues1}
\begin{array}{cl}
&\lambda_i^2(P) \leqslant \MU_i^2(\Sigma)^+ \leqslant \lambda_i^2(P+M) \\
\mbox{ and hence }  &\lambda_i(P) \leqslant \MU_i(\Sigma)^+ \leqslant \lambda_i(P+M) ~.
\end{array}
\end{equation}

Next, note that the system-poles and system-zeros are given by
the eigenvalues sets $-\lambda(P)$ and $-\lambda(P+M)$ respectively. However,
while indeed $p(\Sigma)=-\lambda(P)$, 
the indexing convention followed in Section~\ref{subsec:ordering convention},
would have reversal of element-wise inequalities, and thus 
  from Eqn.~\eqref{eqn:P(P+M) eigenvalues1} we get:
\begin{equation}\label{eqn:P(P+M) eigenvalues2}
z_i(\Sigma) \leqslant \MU_i(\Sigma)^- \leqslant p_i(\Sigma)  \text{ for each }\; i=1,2,\ldots,n.
\end{equation}
This completes proof of Theorem~\ref{thm:SZ betwen Pole Zero pair}.\\

In order to prove Theorem~\ref{thm:SZ with ZIP MIMO}, we use the PBH
test of controllability, 
and using symmetry of $A$, we get that for every eigenvector $x$
of $A$, $B^Tx \neq 0$. Now $P$ has the same set of orthogonal eigenvectors of $A$, therefore for every eigenvector $x$ of $P$ we get that $Mx \neq 0$. 
Therefore utilizing the Statement~4 of Lemma~\ref{lem:product P(P+M)} and Eqn.~\eqref{eqn:P(P+M) eigenvalues2}
we get:
  \begin{equation}\label{eqn:P(P+M) eigenvalues3}
  z_i(\Sigma) < \MU_i(\Sigma)^- < p_i(\Sigma)  \text{ for each }\; i=1,2,\ldots,n.
  \end{equation}
  Further, the system $\Sigma$ exhibits ZIP property then:
  \begin{equation}\label{eqn:ZIP 1}
  z_1\;<p_1\;<\;z_2\;<\;p_2\;<\;z_3<\ldots\;<\; p_{n-1}\;<\;z_n\;<\;p_n~.
  \end{equation}
     Therefore, combining Eqns.~\eqref{eqn:ZIP 1} and~\eqref{eqn:P(P+M) eigenvalues3} we get:
  \begin{equation*}
   z_1<\MU_1<p_1<z_2<\MU_2<p_2<z_3<\ldots< p_{n-1}<z_n<\MU_n<p_n~.
  \end{equation*}  
This completes the proof of Theorem~\ref{thm:SZ with ZIP MIMO} also.
\end{proof}

We saw earlier in Eqn.~\eqref{eqn:P(P+M) eigenvalues2} about how the poles and zeros need not
be interlaced for MIMO systems: an extreme case being when two SISO systems are decoupled subsystems of a MIMO system.
It is interesting to note that for any MIMO system in symmetric state-space  realization,
\emph{irrespective of ZIP property}, each spectral zero
lies between a pole-zero pair, i.e., after
appropriate ordering,  $ z_k(\Sigma_Z)<\MU_k(\Sigma_Z)^-<p_k(\Sigma_Z)$.

\section{Examples} \label{sec:examples}
\noindent
We illustrate the above presented theorems with two examples.
\begin{example} \label{ex:for-scaling}
(Decoupled subsystems:) We consider a MIMO transfer function matrix in which we have two subsystems that
are `decoupled', and we see how a sufficiently high value of
the scaling parameter $\eta$ causes the interlacing of, not just the poles and system-zeros, but also the spectral zeros: like
the SISO case.
\begin{align*}
G=\begin{bmatrix}
1+\frac{1}{s+3}+\frac{1}{s+7} &&0 \\ 0 && 1+\frac{1}{s+4}+\frac{1}{s+8}
\end{bmatrix}~.
\end{align*}
Using Gilbert's state-space realization:
\begin{equation*}
A=\begin{bmatrix}
-3 &0 &0 &0\\0 &-4 &0 &0 \\0 &0 &-7 &0\\0 &0 &0 &-8
\end{bmatrix};\;\;
B=\begin{bmatrix}
1 &0\\0 &1\\1 &0\\0 &1
\end{bmatrix}=C^T;\;\; D=\begin{bmatrix}
1 &0\\0 &1
\end{bmatrix} ~.
\end{equation*}
Therefore the poles, zeros and spectral zeros of the system are:
\begin{equation*}
\begin{array}{ll}
\text{System-poles} \!\!&\!\!: p_1=-8.0,\;p_2=-7.0,\;p_3=-4.0,\;p_4=-3.0,\\
\text{System-zeros}\!\!&\!\!:z_1=-9.2,\;z_2-8.2,\;z_3=-4.8,\;z_4=-3.8,\\
\text{Spectral-zeros}\!\!&\!\!: \MU_1=\pm8.5,\;\MU_2=\pm7.5,\;\MU_3=\pm4.4,\;\MU_4=\pm3.4 ~.
\end{array}
\end{equation*}
The system-poles and system-zeros are not interlaced ($z_1$ and $z_2$ are smaller than $p_1$) due to the choice of the two decoupled SISO subsystems. In order to see the effect of the scaling of feed-through matrix $D$, 
we choose a scalar scaling factor $\eta \in \R_+$\ and define $D=\eta \times \begin{bmatrix}1 &0\\0 &1\end{bmatrix}$.
We increase the scaling factor $\eta$ and tabulate its effect on the system-poles and system-zeros interlacing. 
\end{example}

\newcommand{\nsp}{\hspace*{-1.5mm}} 
\newcommand{\Upsp}{$\vphantom{8^{8^{o}}}$} 
\newcommand{\UPsp}{$\vphantom{8^{8^{8^{8{^8}}}}}$} 

 \begin{table}[H]
   \begin{center}
     \caption{\small Effect of scaling parameter $\eta$ on system-zeros/poles and spectral zeros interlacing: of Example~\ref{ex:for-scaling} } \label{tab:table1} 
  \Scale[0.9]{   \begin{tabular}[c]{@{}|p{2.1cm}|p{2.2cm}|rrrr|@{}} 
       \hline
       \footnotesize Scaling parameter $\eta$, remark  & \footnotesize System pole/zero properties &\multicolumn{1}{c}{(1)} &\multicolumn{1}{c}{(2)} &\multicolumn{1}{c}{(3)} &\multicolumn{1}{c|}{(4)}\\ \hline 
       \multirow{3}{*}{
         \Scale[0.92]{ \nsp \begin{tabular}{c}
           $\eta=1$\\system-poles/zeros: \\
           not interlaced 
       \end{tabular} } }
       &system-zeros $(z_i)$ &\textbf{-9.24}\Upsp  &\textbf{-8.24 }  &-4.76   &-3.76\\
       &spectral-zeros $(\MU_i)$ & -8.52   &-7.51   &-4.40   &-3.40\\
       &system-poles $(p_i) $ &\textbf{-8.00}  &-7.00  &-4.00   &-3.00 \\
       \hline
       \multirow{3}{*}{
         \Scale[0.92]{  \nsp\begin{tabular}{c}
           $\eta=1.2$\\ system-poles/zeros: \\
           verge-of interlaced 
       \end{tabular} }  }
       &system-zeros $(z_i)$ &\textbf{-9.00}\Upsp  &\textbf{-8.00}   &-4.67   &-3.67\\
       &spectral-zeros $(\MU_i)$ & -8.43   &-7.43   &-4.35   &-3.34\\
       &system-poles $(p_i) $ &\textbf{-8.00}  &-7.00  &-4.00   &-3.00 \\ \hline
       \multirow{3}{*}{
       \Scale[0.92]{  \nsp  \begin{tabular}{c}
           $\eta=2$\\ system-poles/zeros:\\
           after interlaced 
       \end{tabular} }   }
       &system-zeros $(z_i)$ &-8.56\Upsp  &-7.56   &-4.44   &-3.44\\ 
       &spectral-zeros $(\MU_i)$ & -8.26   &-7.25   &-4.22   &-3.22\\
       &system-poles $(p_i) $ &-8.00  &-7.00  &-4.00   &-3.00\\ \hline
       \multirow{4}{*}{ \vspace*{10mm}
        \Scale[0.90]{  \nsp \begin{tabular}{c}
           $\eta=100$\\  system-poles/zeros:\\ 
           continue to be\\
   interlaced
       \end{tabular} } }
       &system-zeros \UPsp $(z_i)$ &-8.01\Upsp  &-7.01   &-4.01   &-3.01\\ 
       &spectral-zeros $(\MU_i)$ & -8.005   &-7.005   &-4.005   &-3.005\\
       & system-poles $(p_i) $ & -8.00  &-7.00  &-4.00   &-3.00\\[3mm]   \hline
     \end{tabular} }
   \end{center} 
 \end{table} 

From Table~\ref{tab:table1}, it is evident that that as the scaling factor $(\eta)$ is increased system-zeros and system-poles move closer to interlace condition.
When the feed-through matrix $D$ becomes \textit{sufficiently large}  i.e. $\eta\geqslant 2$ then the system-poles and system-zeros along with spectral zeros get  interlaced. 

\begin{example} \label{def:example-multiagent}
(Multi-agent example:) Consider a multi-agent network arranged in a path graph as shown in the figure below:
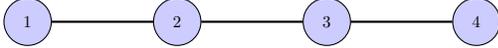
\begin{figure}[h]
  \centering
  \begin{tikzpicture}[scale=0.9,transform shape, every node/.style={scale=0.7}]
  \node[main node] (1) {$1$};
  \node[main node] (2) [right = 1.5cm of 1]  {$2$};
  \node[main node] (3) [right = 1.5cm of 2]  {$3$};
  \node[main node] (4) [right = 1.5cm of 3]  {$4$};
  
  \path[draw,thick]
  (1) edge node {} (2)
  (2) edge node {} (3)
  (3) edge node {} (4);
  \end{tikzpicture}
  \caption{\small Multi-agent network in path graph arrangement}
\end{figure}
If we  consider that the node-$1$ and node-$4$ are chosen as the controlled nodes and inputs to the nodes are current $i_1$ and $i_4$ injected in the node. 
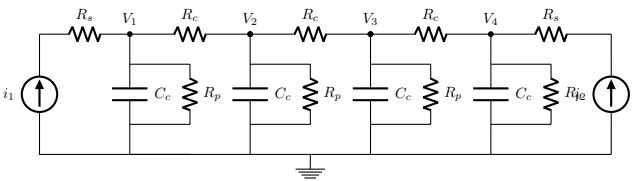
\begin{figure}[!h]
  \centering
  \ctikzset{bipoles/resistor/height=0.25}
  \ctikzset{bipoles/resistor/width=0.5}
  \begin{tikzpicture}[scale=0.8,transform shape, every node/.style={scale=0.7}]
  \draw (1.5,0) to[R=$R_s$]  (3,0) to[R=$R_c$, *-*] (5,0) to[R=$R_c$,*-*] (7,0) to[R=$R_c$, *-*] (9,0) to[R=$R_s$]  (11,0);
  \draw (4,-2) -- (1.5,-2) to[I=$i_1$] (1.5,0);
  \draw (9,-2) -- (11,-2) to[I=$i_2$] (11,0)  ;
  \draw (3,0) to[C=$C_c$] (3,-2);
  \draw (3,-0.5) -- (4,-0.5) to[R=$R_p$] (4,-1.5) -- (3,-1.5);
  \draw (5,0) to[C=$C_c$] (5,-2) -- (3,-2);
  \draw (5,-0.5) -- (6,-0.5) to[R=$R_p$] (6,-1.5) -- (5,-1.5);
  \draw (7,0) to[C=$C_c$] (7,-2) -- (5,-2);
  \draw (7,-0.5) -- (8,-0.5) to[R=$R_p$] (8,-1.5) -- (7,-1.5);
  \draw (9,0) to[C=$C_c$] (9,-2) -- (7,-2);
  \draw (9,-0.5) -- (10,-0.5) to[R=$R_p$] (10,-1.5) -- (9,-1.5);
  \draw (6,-2) node[ground]{}; 
  \node[] at (3,0.25) {$V_1$}; 
  \node[] at (5,0.25) {$V_2$}; 
  \node[] at (7,0.25) {$V_3$}; 
  \node[] at (9,0.25) {$V_4$};
  \ctikzset{resistor = american}
  \end{tikzpicture}
  \caption{\small Multi-agent network with sources at two nodes}
  \label{fig:Network}
\end{figure}

If the node voltages $(V_1,V_2,V_3,V_4)$ are considered as the state variables and the inputs are $i_1$ and $i_4$ with outputs as the voltages across the current sources then, the state-space realization can be expressed as:
\begin{equation*}\Scale[0.94]{
\! \! \begin{array}{l}
\! \!   A \!  =\!  \begin{bmatrix}
    \frac{-1}{R_c C_c}-\frac{1}{R_pC_c} &\frac{1}{R_c C_c} &0 &0 \\ 
    \frac{1}{R_c C_c} &\frac{-2}{RC_c}-\frac{1}{R_pC_c} &\frac{1}{R_c C_c} &0 \\
   0 &\frac{1}{R_c C_c} & \frac{-2}{RC_c}-\frac{1}{R_pC_c} &\frac{1}{R_c C_c} \\
   0 & 0 &\frac{1}{R_c C_c} & \frac{-1}{RC_c}-\frac{1}{R_pC_c} \end{bmatrix} \\
\!\!    B \! =\! \begin{bmatrix}\frac{1}{C_c} &0\\0 &0\\0 &0\\0 &\frac{1}{C_c} \end{bmatrix}, 
  C =\begin{bmatrix}1 &0 &0 &0\\ 0 &0 &0 &1\end{bmatrix}, \mbox{ and }
  D =\begin{bmatrix}
  R_s &0\\0 &R_s 
  \end{bmatrix}~.
  \end{array}
}\end{equation*}

\noindent
Assuming numerical values: $ R_s=0.5~\Omega \; R_c=1~\Omega, \; C_c=1~F,\; R_p=10~\Omega$ we get: 
\begin{equation*}\Scale[0.9]{
A= \begin{bmatrix}
 -1.1  & 1.0   & 0.0   & 0.0\\  
 1.0    &-2.1   &1.0   & 0.0\\  
 0.0    & 1.0   &-2.1   &1.0\\  
 0.0    &0.0    &1.0   &-1.1
 \end{bmatrix},\;
 B\! =\! C^T=\! \begin{bmatrix}
 1  & 0\\ 
 0  &0 \\ 
 0  &0\\ 
 0  &1
 \end{bmatrix},\;
 D\! =\! 0.1\cdot \begin{bmatrix}
\! 1\!  &\! 0 \! \\ \! 0 \! &\! 1 \!
 \end{bmatrix}.
}\end{equation*}

\noindent
Therefore, the poles and zeros of the systems are:
\begin{equation*}
\Scale[0.94]{ \begin{array}{l}
\text{System-poles}:\;\!  p_1=-3.51,\;p_2=-2.10,\;p_3=-0.69,\;p_4=-0.10, \\
\text{System-zeros}:\;\!  z_1=-11.22,\;z_2=-11.20,\;z_3= -2.98,\;z_4=-1.0~.
\end{array} }
\end{equation*}
We can see that the poles and zeros are not interlaced ($z_1$ and $z_2$ both
are less than $p_1$).
Now to see the effect of the scaling of feed-through matrix $D$, 
we choose a scalar scaling factor $\eta \in \R_+$\ and define
$D:=\eta \times \begin{bmatrix}0.1 & 0\\ 0 & 0.1\end{bmatrix}$.
\end{example}
We increase the scaling factor and tabulate its effect on the system-poles and system-zeros interlacing. 
\begin{table}[H]
\begin{center}
\caption{\small Effect of scaling parameter $\eta$ on system-zeros/poles and spectral zeros interlacing: Example~\ref{def:example-multiagent}} \label{tab:table2}
\Scale[0.92]{ \begin{tabular}[c]{@{}|p{2.1cm}|p{2.2cm}|rrrr|@{}} 
  \hline
Scaling parameter $\eta$, remark & System pole/zero properties &\multicolumn{1}{c}{(1)} &\multicolumn{1}{c}{(2)} &\multicolumn{1}{c}{(3)} &\multicolumn{1}{c|}{(4)}\\ \hline 
\multirow{3}{*}{
\Scale[0.92]{ \nsp \begin{tabular}{c}
$\eta=1$,\\system-poles/zeros:\\
not interlaced 
\end{tabular} } }
  & system-zeros $(z_i)$ &\textbf{-11.22}\Upsp  &\textbf{-11.20}   &-2.98   &-1.00\\
  &  spectral-zeros $(\MU_i)$ & -4.44   &-3.91   &-2.02   &-0.39\\
 & system-poles $(p_i) $ &\textbf{-3.51}  &-2.10  &-0.69   &-0.10 \\
 \hline
\multirow{3}{*}{
\Scale[0.92]{ \nsp \begin{tabular}{c}
 $\eta=5$,\\ system-poles/zeros:\\
verge-of interlaced 
\end{tabular} }  }
&   system-zeros $(z_i)$ &\textbf{-4.10}\Upsp  &\textbf{-3.51}   &-2.10   &-0.69\\
  &  spectral-zeros $(\MU_i)$ & -3.67   &-2.56   &-1.24   &-0.28\\
 &  system-poles $(p_i) $ & \textbf{-3.51}  &-2.10  &-0.69   &-0.10 \\ \hline
\multirow{3}{*}{
\Scale[0.92]{ \nsp  \begin{tabular}{c}
$\eta=10$,\\system-poles/zeros:\\
after interlaced \end{tabular} }
}  &   system-zeros $(z_i)$ &-3.72\Upsp  &-2.72   &-1.48   &-0.48\\ 
 & spectral-zeros $(\MU_i)$ & -3.59   &-2.34   &-1.01   &-0.22\\
 &system-poles $(p_i) $ & -3.51  &-2.10  &-0.69   &-0.10\\ \hline
\multirow{4}{*}{ \vspace*{10mm}
\Scale[0.90]{ \nsp  \begin{tabular}{c}
$\eta=100$,\\ system-poles/zeros:\\
continue to be\\
 interlaced
\end{tabular} } }
  &
system-zeros \UPsp $(z_i)$ &-3.53\Upsp  &-2.15   &-0.77   &-0.15\\ 
 & spectral-zeros $(\MU_i)$ & -3.52   &-2.12   &-0.73   &-0.12\\
 &system-poles $(p_i) $ &-3.51  &-2.10  &-0.69   &-0.10 \\[3mm]   \hline
\end{tabular} }
\end{center}
\end{table} 
From Table~\ref{tab:table2}, we infer that as the scaling factor $(\eta)$ is increased system-zeros and system-poles move closer to interlace condition. 
When the feed-through matrix $D$ is \textit{sufficiently large}
i.e. $\eta \geqslant 10$, then the system-poles and system-zeros together
with spectral zeros get interlaced. 

\section{Concluding remarks} \label{sec:conclusion}
We first studied balancing of strictly passive systems using extremal
solutions of its Algebraic Riccati Equation (ARE). 
The extremal ARE solutions $\Kmin$ and $\Kmin$ induce quadratic functions
that signify the energy available and required supply for a given state $x$
with respect to the passivity supply rate $u^Ty$. While
positive real balancing aims to have $\Kmax = \Kmin^{-1}$,
we introduced other forms of quasi-balancing and showed inter-relations
between these realizations and between the positive real singular values.
We also proved the relevance in the context
of systems admitting a symmetric state-space realization: these systems played
a central role in this paper, especially for MIMO systems' ZIP properties.

In Section~\ref{sec:siso:interlacing} we first studied
the properties of spectral-zero for strictly passive SISO systems: spectra-zeros'
realness and their interlacing with
system poles/zeros (Theorem~\ref{thm:real:spectral:zeros_siso}).
We also proved in Lemma~\ref{lem:sum of squares} the relation between
the product and sum of squares of the spectral zeros with
the system-poles and system-zeros, and obtained as a special case
that for an order-$1$ system, the spectral-zero is
the geometric mean of the system-pole and system-zero.

In the context of MIMO systems, we first used the 
definition of system-zeros of a system, say $\Sigma_Z$, as
the poles of its inverse system $\Sigma_Y$, and proved that both $\Sigma_Z$ and
$\Sigma_Y$ have the same Hamiltonian matrix and the same spectral zeros; a property
specific to the i/o invariant supply rate: $u^T y$. Next, pursuing
with systems that admit a 
symmetric state-space realization, we next proved realness of all the spectral-zeros
(Theorem~\ref{thm:real:spectral:zeros_mimo}). 
Using existing/new properties of differences in eigenvalues of
pairs of symmetric matrices, we proved in Theorem~\ref{thm:ZIP MIMO} that if
the poles of a MIMO
are `relatively well-separated', then the poles and zeros are interlaced.
We also showed that this separation is ensured by systems with a sufficiently
large feed-through matrix (Lemmas~\ref{lem:sufficiently-large-D-for-Z} and \ref{lem:sufficiently-large-D-for-Y}).
Finally, we prove that strictly passive
systems with symmetric state-space realizations allow
not just ZIP but also spectral zero interlacing: Theorems~\ref{thm:SZ with ZIP MIMO}
and~\ref{thm:SZ betwen Pole Zero pair}.

In Section~\ref{sec:examples}, we elaborated on
a few examples (linked to the RC/RL network of Section~\ref{subsec:network-example},
and a multi-agent network), for which the results in our paper were applicable.

A possible direction for further work is to formulate milder or other
sufficient conditions that guarantee realness of spectral zeros and/or
interlacing properties of system poles/zeros: it is possible that
under a different definition of system-zero, conditions
for ZIP for MIMO systems would be different and also both necessary and
sufficient. Further, a relationship between the positive real singular values and positive real quasi-singular values with the system-poles/zeros and spectral-zeros can be derived for systems with ZIP so that they can be computed without solving the ARE.

Another direction of further research is to explore the extent to which
Model-Order-Reduction methods developed for ZIP SISO systems are extendable
to symmetric state-space MIMO systems having the ZIP property.



%

\bibliographystyle{IEEEtran}

\vspace*{3mm}

\noindent
\includegraphics[width=0.17\columnwidth]{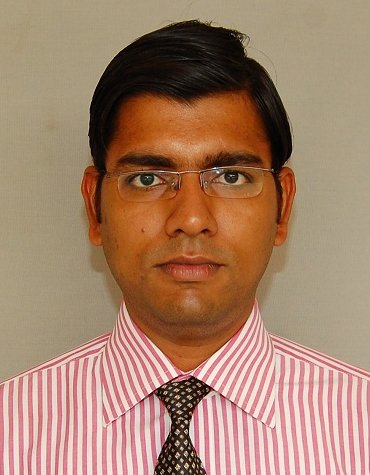}
~~\parbox[b]{0.76\columnwidth}{\small \textbf{Sandeep Kumar} received his Bachelors
        degree in Electrical Engineering from Sardar Vallabhbhai National
        Institute of Technology, Surat, India in 2001.
        He is currently a Scientist in Advanced Centre for Energetic
        Materials, DRDO, Nashik, India and also is pursuing Ph.D. at IIT Bombay.}

\vspace*{3mm}
\noindent
\includegraphics[width=0.17\columnwidth]{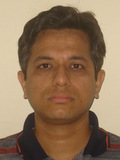}
~~\parbox[b]{0.76\columnwidth}{\small \textbf{Madhu N. Belur} is at IIT Bombay since 2003,
        where he currently is a professor in the Department of Electrical Engineering.
        His interests include dissipative dynamical systems, graph theory and
        open-source implementation for various applications.}

\end{document}